\documentclass{LMCS}

\def\dOi{11(1:19)2015}
\lmcsheading%
{\dOi}
{1--26}
{}
{}
{Jan.~20, 2014}
{Mar.~31, 2015}
{}

\ACMCCS{[{\bf Theory of computation}]: Logic---Finite Model Theory;
  Formal languages and automata theory---Tree languages}


\usepackage{amsmath,amssymb,bbm,mathrsfs,graphicx,euscript}
\usepackage{verbatim}
\usepackage{hyperref}
\usepackage{tikz}
\tikzstyle{every picture} = [>=latex]
\usepackage{xspace}

\def\ca#1{{\mathcal#1}}
\def\cf#1{{\EuScript#1}}
\let\sem\setminus

\def\prebox#1{\mathop{\mbox{\sl#1}}}

\newcommand\cwd{\operatorname{\mathit{cw}}}
\newcommand\tw{\operatorname{\mathit{tw}}}
\newcommand\td{\operatorname{\mathit{td}}}

\newcommand\towf[1]{\operatorname{\mathit{exp}}^{(#1)}}

\newcommand{\FO}{\ensuremath{\mathrm{FO}}\xspace}
\newcommand{\LinEMSO}{\ensuremath{\mathrm{LinEMSO}}\xspace}
\newcommand{\MSO}{\ensuremath{\mathrm{MSO}}\xspace}
\newcommand{\CMSO}{\ensuremath{\mathrm{CMSO}}\xspace}
\newcommand{\MSOi}{\ensuremath{\mathrm{MSO}_1}\xspace}
\newcommand{\CMSOi}{\ensuremath{\mathrm{CMSO}_1}\xspace}
\newcommand{\CMSOii}{\ensuremath{\mathrm{CMSO}_2}\xspace}
\newcommand{\MSOii}{\ensuremath{\mathrm{MSO}_2}\xspace}

\def\modp#1#2{\prebox{mod}_{#1,#2}}

\def\Nrec#1#2#3#4{N_{#2,#3,#4}(#1)}
\def\Trecf#1#2#3{R_{#1,#2,#3}}
\def\Trec#1#2#3#4{\Trecf{#2}{#3}{#4}(#1)}
\def\TrecfM#1#2#3{\Trecf{#1}{#2}{#3}^{*M}}
\def\TrecM#1#2#3#4{\TrecfM{#2}{#3}{#4}(#1)}
\def\TM#1#2{{\mathcal T\!M}_{#2}(#1)}

\def\pconsistent#1{\prebox{consistent}\nolimits_{\,#1}}
\def\pseparable#1{\prebox{separable}\nolimits_{\,#1}}
\def\preducex(#1,#2){\prebox{reduce}\nolimits_{\,#1,#2}}
\def\preduceto(#1){\prebox{reduce-to}\nolimits_{\,#1}}
\def\pconnected{\prebox{connected}\,}

\begin{document}
\title[Kernelizing MSO Properties of Trees of Fixed Height]{Kernelizing MSO Properties of Trees of Fixed Height, and Some Consequences\rsuper*}
\titlecomment{{\lsuper*}Part of the results have been published in the FSTTCS 2012
Conference, Dagstuhl LIPIcs Series, 2012}	

%

\author[J.~Gajarsk\'y]{Jakub~Gajarsk\'y}	
\address{Faculty of Informatics, Masaryk University, Botanick\'a 68a, Brno, Czech Republic}	
\email{\{gajarsky,hlineny\}@fi.muni.cz}  

\author[P.~Hlin\v{e}n\'y]{Petr~Hlin\v{e}n\'y}	
\address{\vspace{-18 pt}}	
\thanks{Both authors have been supported by the Czech Science Foundation; project no.~14-03501S}	

\begin{abstract}
Fix an integer $h\geq1$.
In the universe of coloured trees of height at most $h$,
we prove that for any graph decision problem defined by
an \MSOi formula with $r$ quantifiers, 
there exists a set of kernels, each of size bounded
by an elementary function of $r$ and the number of colours. 
This yields two noteworthy consequences.
Consider any graph class $\cf G$ having a one-dimensional \MSOi interpretation
in the universe of coloured trees of height $h$
(equivalently, $\cf G$ being a class of shrub-depth~$h$).
First, $\cf G$ admits an \MSOi model checking algorithm
whose runtime has an elementary dependence on the formula size.
Second, on~$\cf G$ the expressive powers of \FO and \MSOi coincide
(which extends a 2012 result of Elberfeld, Grohe, and Tantau).

\end{abstract}

\maketitle

\section{Introduction}

First order (\FO) and monadic second-order (\MSO) logics play an undoubtedly
crucial role in computer science.
Besides traditional tight relations to finite automata and regular
languages, this is also witnessed by their frequent occurrence in the 
so called {\em algorithmic metatheorems}
which have gained increasing popularity in the past few years.
The term algorithmic metatheorem commonly refers to a general algorithmic 
toolbox ready to be applied onto a wide range of problems in specific
situations, and \MSO or \FO logic is often used in the expression of 
this ``range of problems''.

One of the perhaps most celebrated algorithmic metatheorems (and the
original motivation for our research)
is Courcelle's theorem \cite{cou90} 
stating that every {graph property} $\phi$ expressible in the
{\em\MSOii logic of graphs} (allowing for both vertex and edge set quantifiers)
can be decided in linear fpt time on graphs of bounded tree-width.
Courcelle, Makowsky, and Rotics \cite{cmr00} then have analogously addressed 
a wider class of graphs, namely those of bounded clique-width,
at the expense of restricting $\phi$ to
{\em\MSOi logic} (i.e., with only vertex set quantification).

Regarding Courcelle's theorem \cite{cou90} and closely related
\cite{als91,cmr00}, it is worth to remark that a solution can be
obtained via translating of the respective graph problem
to an \MSO formula over coloured trees
(which relates the topic all the way back to Rabin's S2S theorem~\cite{Rab69} and works of Doner~\cite{Doner} and Thatcher and Wright~\cite{Thatcher_Wright}).
However, a common drawback of these metatheorems is that,
when their runtime is expressed as \mbox{$\ca O\big(f(\phi,width(G))\cdot|G|\big)$},
this function $f$ grows asymptotically as 
$2^{\left.2^{.^{.^{width(G)}}}\right\}a}$ where the height $a$ depends on $\phi$,
precisely on the quantifier alternation depth of~$\phi$
(i.e., $f$ is a {\em non-elementary function} of the parameter~$\phi$).
The latter is not surprising since Frick and Grohe \cite{FG04} 
proved that it is not possible to avoid a non-elementary
tower of exponents in deciding \MSO properties on all trees
or coloured paths (unless P=NP),
and Lampis \cite{Lampis-ICALP13} proved an analogous negative result 
even for uncoloured paths (unless EXP=NEXP).

The aforementioned negative results leave room for possible improvement
on suitably restricted subclass(es) of all coloured trees,
namely on those avoiding long paths.
In this respect, our first result
(Theorem~\ref{thm:dtreekern}) 
gives a new algorithm for deciding \MSO properties $\phi$ of 
rooted coloured trees $T$ of fixed height~$h$.
The algorithm (Corollary~\ref{cor:dtreealg}) uses so called kernelization---which 
means it efficiently reduces the input tree into an equivalent one 
(the {\em kernel}) of elementarily bounded size; by
$$ 2^{2^{\left..^{.^{\ca O(|\phi|^2)}}\right\}h+1}}.$$

In the complexity aspects, our result ``trades'' quantifier alternation
depth of $\phi$ from Courcelle's theorem for bounded height of the tree.
Again, the tower of exponents of (this time fixed) height $h+1$ in the expression
is unavoidable unless the Exponential Time Hypothesis fails, 
as proved by aforementioned Lampis~\cite{Lampis-ICALP13}.
We refer to Section~\ref{sec:reductionlem}
for an exact expression of runtime
as well as for an extension to counting \MSO logic.

From a more general perspective our algorithm can be straightforwardly applied to
any suitable ``depth-structured'' graph class via efficient
interpretability of logic theories, such as to graph classes of bounded
tree-depth or of bounded shrub-depth~\cite{GHNOOR12}.
This (asymptotically) includes previous results of Lampis~\cite{Lam12} and 
Ganian~\cite{Gan11} as special cases.
Even more, the scope of our result can be extended to the so called \LinEMSO 
optimization and enumeration framework, see e.g.~in \cite{cmr00},
over such graph classes as follows:
the algorithmic metatheorems of \cite{cou90,als91,cmr00} (and similar ones) can be treated
using finite tree-automata (with non-elementary numbers of states in general),
however, in the universe of a suitable ``depth-structured'' graph class
only very few of the automaton states correspond to some of the kernels
and so are actually reachable.
This in Section~\ref{sec:algoEMSO} concludes the first half of our paper.

\smallskip

There are also other sides of the main result.
First, the initial discovery of Theorem~\ref{thm:dtreekern} was the prime
motivation for defining shrub-depth in~\cite{GHNOOR12}, and a key ingredient
in the proof that shrub-depth is stable under \MSOi interpretations,
again in~\cite{GHNOOR12}. 

Second, Elberfeld, Grohe, and Tantau
\cite{EGT12} prove that \FO and \MSOii have equal expressive power on
the graphs of bounded tree-depth.
Having Theorem~\ref{thm:dtreekern} at hand, we can provide a relatively
simple alternative proof of this result.
Furthermore, using some more sophisticated combinatorial tools, namely
well-quasi-ordering, we prove a new result (Theorem~\ref{thm:FOMSOnew}) that 
\FO and \MSOi have equal expressive power on any graph class of bounded shrub-depth.
In the converse direction, Elberfeld, Grohe, and Tantau
\cite{EGT12} also prove that on monotone graph classes of unbounded
tree-depth, \MSOii is strictly stronger than \FO.
Unfortunately, due to lack of a suitable ``forbidden substructure''
characterization of shrub-depth, we are not yet able to prove the analogous converse
claim, but we conjecture that a hereditary class on which \FO and \MSOi
coincide must have bounded shrub-depth (Conjecture~\ref{conj:samepower-shrub}).

\section{Preliminaries}
\label{sec:def}

We assume standard terminology and notation of graph theory,
see e.g.\ Diestel~\cite{die05}. 
Our graphs are finite, simple and undirected by default.
When dealing with trees, we implicitly consider them as rooted
(with the implicit parent-child tree order) and with unordered descendants.
We use $\subseteq$ for the usual subgraph relation and $\subseteq_i$ for
induced subgraphs.
A graph class  $\cf C$ is \emph{hereditary} if it is closed under the induced 
subgraph relation, i.e.,
$G\in\cf C$ and $H\subseteq_i G$ implies $H\in\cf C$.

A class $\cf C$ of graphs is {\em well-quasi-ordered} (WQO) under an order $\preceq$
if, for any infinite sequence $(G_1,G_2,\dots)\subseteq\cf C$,
it is $G_i\preceq G_j$ for some $i<j$. 
WQO techniques are very popular in structural graph theory, and we refer
to \cite[Chapter~12]{die05} for a brief overview.
A graph property $\Pi$ is {\em preserved under $\preceq$} if the following holds;
whenever $G \in \cf C$ has the property $\Pi$, 
every $G' \in \cf C$ such that $G' \preceq G$ has $\Pi$ as well.
The following simple folklore claim is crucial in this paper:
\begin{prop}
\label{prop:obstacles}
Let $\cf C$ be a graph class.
If $\Pi$ is a property preserved under a well-quasi-order $\preceq$ on $\cf C$,
then there exists a finite set $Obst\subseteq\cf C$ such that;
$G\in\cf C$ has $\Pi$ if and only if $H\not\preceq G$ for all $H\in Obst$.
\qed
\end{prop}

We informally say that $\Pi$ from Proposition~\ref{prop:obstacles} has finitely
many obstacles in~$\cf C$.
We are going to use the following well-quasi-ordered graph class.
\begin{thm}[Ding~\cite{Ding92}]
\label{thm:Ding}
Let $m\in\mathbb N$ be an integer and $C$ be a finite set of colours.
The class of the graphs not containing a path on $m$ vertices as a
subgraph and with vertices coloured by $C$ is well-quasi-ordered
under the colour-preserving induced subgraph order $\subseteq_i$.
\qed
\end{thm}

Given a graph~$G$, a \emph{tree-decomposition of $G$} is an ordered pair
$(T,\mathcal{W})$, where~$T$ is a tree and 
$\mathcal{W} = \{W_x \subseteq V(G) \mid x \in V(T)\}$ is a collection
of {\em bags} (vertex sets of~$G$), 
such that the following hold:
\begin{enumerate}
    \item $\bigcup_{x \in V(T)} W_x = V(G)$;
    \item for every edge~$e = uv$ in~$G$, there exists~$x \in V(T)$ such that~$u,v \in W_x$;
    \item for each~$u \in V(G)$, the set~$\{x \in V(T) \mid u \in W_x\}$
    induces a subtree of $T$.
\end{enumerate}
The {width} of a tree-decomposition $(T,\mathcal{W})$ is 
$(\max_{x \in V(T)}|W_x|)-1$.
The \emph{tree-width} of~$G$, denoted~$\tw(G)$, is the smallest width of a
tree-decomposition of~$G$. 

Besides tree-width, another useful width measure of graphs is the
{\em clique-width} of a graph $G$.
This is defined for a graph $G$ as the smallest number of labels $k=\cwd(G)$
such that some labelling of $G$ can be defined
by an algebraic {\em k-expression} using the following operations:
\begin{enumerate}
\item create a new vertex with label $i$;
\item take the disjoint union of two labelled graphs;
\item add all edges between vertices of label $i$ and label $j$; and
\item relabel all vertices with label $i$ to have label $j$.
\end{enumerate}

\smallskip
{\em Monadic second-order logic} ($\MSO$)
is an extension of first-order logic (\FO) by quantification over sets. 
The \emph{quantifier rank} $qr(\phi)$ is the nesting depth of quantifiers in $\phi$.  
The formulas of quantifier rank $0$ are called \emph{quantifier free}.
{\em Counting monadic second-order logic} ($\CMSO$) is an extension of
$\MSO$ which allows use of predicates $\prebox{mod}_{a, b}(X)$, where $X$
is a set variable.  The semantics of the predicate $\prebox{mod}_{a, b}(X)$ is
that the set $X$ has $a$ modulo $b$ elements.
 
Let $\sigma$ and $\tau$ be relational vocabularies and let $L \in\{\FO,
\MSO,\CMSO\}$. 
A~\emph{one-dimensional interpretation}%
\footnote{The name \emph{one-dimensional (noncopying) transduction} is also used 
in an algorithmic context, see, e.g.,~\cite{CE12}.
Transductions mean, however, usually a more general concept than what we 
define and use here.}
of $\tau$ in $\sigma$ is a tuple
$I = \big(\nu(x),\, \{\eta_R(\bar{x})\}_{R\in \tau}\big)$ of $L[\sigma]$-for\-mulas where
$\nu$ has one free variable and the number of free variables in each
$\eta_R$ is equal to the arity of $R$ in $\tau$.
\begin{itemize}
\item
To every $\sigma$-structure $A$ the interpretation $I$ assigns a $\tau$-structure
$A^I$ with the domain $A^I = \{a~|~A \models \nu(a) \}$ and the relations $R^I =\{
{\bar{a}~|~A \models \eta_R(\bar{a})}\}$ for each $R \in \tau$.  We say that
a class $\cf{C}$ of $\tau$-structures has an interpretation in a class
$\cf{D}$ of $\sigma$-structures if there exists an interpretation $I$
such that for each $C \in \cf{C}$ there exists $D \in \cf{D}$ such
that $C \simeq D^I$, and for every  $D \in \cf{D}$ the structure $D^I$
is isomorphic to a member of~$\cf C$.
\item
The interpretation $I$ of $\tau$ in $\sigma$ defines a translation of every
$L[\tau]$-formula $\psi$ to an $L[\sigma]$-formula $\psi^I$ as follows:
\begin{itemize}
\item every $\exists x. \phi$ is replaced by $\exists x.(\nu(x) \land \phi^I)$,
\item every $\exists X. \phi$ is replaced by 
$\exists X.(\forall y (y \in X \rightarrow \nu(y)) \land \phi^I)$,
and
\item every occurrence of a $\sigma$-atom $R(\bar{x})$ is replaced by
the corresponding formula~$\eta_R(\bar{x})$.
\end{itemize}  
\end{itemize}  
We have added the adjective ``one-dimensional'' (interpretation) to indicate that
our $\nu$ is of arity one, i.e., that the domain of
$\tau$-structures is interpreted in singleton elements of the $\sigma$-structures.
Since we use only one-dimensional interpretations throughout the paper,
from now on
we will say shortly an ``{\em interpretation}'' to mean a one-dimensional interpretation.

We make use of the following claim:
\begin{lem}[\cite{model_th_book}]
\label{lem:interpretation}
Let $I$ be interpretation of $\tau$ in $\sigma$. 
Then for all $L[\tau]$-formulas $\phi$ and all $\sigma$-structures $A$
$$ A \models \phi^I \Longleftrightarrow A^I \models \phi .$$
\qed
\end{lem}

A great part of our paper deals with \MSO logic of rooted trees,
which are structures with a single parent-child binary relation.
For general graphs, however, there are two established but inequivalent views
of them as relational structures.
In the one-sorted adjacency model of graphs, \MSO specifically reads as follows:
\begin{defi}[\MSOi logic of graphs]
\label{df:MSO1}
The language of $\MSOi$ 
contains the expressions built from the following elements:
\begin{itemize}
\item variables $x,y,\ldots$ for vertices, and $X,Y,\ldots$ for sets of vertices,
\item the predicates $x\in X$ and $\prebox{edge}(x, y)$ with the standard meaning,
\item
   equality for variables, 
   the connectives~$\land, \lor, \lnot, \to$, and the
   quantifiers~$\forall, \exists$ over vertex and vertex-set variables.
\end{itemize}
\end{defi}

\noindent $\MSOii$ logic of graphs extends $\MSOi$ by allowing quantification over edge sets. Formally, 
one can consider graphs as two-sorted structures (the two sorts being vertex-set and edge-set of a graph) with adjacency and incidence predicates.

Although our results are concerned also with $\MSOii$, 
we refrain from giving full definition of $\MSOii$ logic of graphs, since by the 
following theorem we can avoid using $\MSOii$ explicitly.

\begin{thm}[Courcelle~\cite{Cou94}, also~\cite{CE12}]
\label{thm:MS1=MS2}
Let $\cf{C}$ be the class of finite graphs of tree-width at most $k$ (for any fixed k). 
A property of graphs in
$\cf{C}$ is $\CMSOii$-expressible if, and only if, it is $\CMSOi$-expressible. 
Moreover, for any fixed $k$, if $\phi$ is an $\CMSOii$ sentence, the size of
an equivalent $\CMSOi$ sentence $\phi'$ can be bounded by an elementary
function of $|\phi|$.%
\footnote{%
  We note that the theorem is explicitly stated only for $\MSOi$ and $\MSOii$
  in ~\cite{Cou94,CE12} (Theorems 1.44 and 5.22 of~\cite{CE12}). 
  To see that the same holds for $\CMSOi$ and $\CMSOii$ see Section~5.2.6 of~\cite{CE12}.
  The second part of the theorem concerning the size of an equivalent
  $\CMSOi$ formula can be obtained from the proofs.
  Therefore, we cannot cite a precise bound of $|\phi'|$ in terms of $|\phi|$,
  but one can read from the proofs that the bound is actually polynomial and
  even linear for a fixed~$k$.
}
\qed
\end{thm}

We often use labelled graphs, i.e.
graphs where each vertex can have labels from some finite set $Lab$. Labels are modelled by unary predicates. Sometimes we refer to
colours instead of labels; the colour of a vertex is the combination of its labels (each vertex has exactly one
colour, the number of colours is $\sim 2^{|Lab|}$).

For an introduction to parameterized complexity we suggest~\cite{df99}.
Here we just recall that a problem $\ca P$ parameterized by $k$, i.e., with an input
$\langle x,k\rangle\in \Sigma^*\times\mathbb N$, is {\em fixed parameter
tractable}, or {\em fpt}, if it admits an algorithm in time
$\ca O\big(f(k)\cdot|x|^{\ca O(1)}\big)$ where $f$ is an arbitrary
computable function.

\section{Trees of Bounded Height and \MSO}
\label{sec:reductionlem}

\let\MSOtree\MSO
\let\CMSOtree\CMSO

The primary purpose of this section is to prove Theorem~\ref{thm:dtreekern};
that for any $m$-coloured tree $T$ of constant height $h$ there exists 
an efficiently computable subtree $T_0\subseteq T$ (a {\em kernel})
such that, for any \MSOtree sentence $\phi$ of fixed quantifier rank $r$, it is
$T\models\phi$ $\iff$ $T_0\models\phi$,
and the size of $T_0$ is bounded by an elementary function of $r$ and $m$ 
(the dependence on $h$ being non-elementary, though).
Particularly, since checking of an \MSOtree property $\phi$ can be easily
solved in time $\ca O^*\big(2^{c|\phi|}\big)$ on a graph with $c$ vertices
(in this case $T_0$) by recursive exhaustive expansion of all quantifiers of
$\phi$, this gives a kernelization-based elementary fpt algorithm for \MSOtree 
model checking of rooted $m$-coloured trees of constant height $h$
(Corollary~\ref{cor:dtreealg}).

We need a bit more formal notation.
The {\em height}\,\footnote{ There is a conflict in the
  literature about whether the height of a rooted tree should be measured by
  the ``root-to-leaves distance'' or by the ``number of levels'' (a
  difference of $1$ on finite trees). We adopt the convention that the
  height of a single-node tree is $0$ (i.e., the former view).} 
$h$ of a rooted tree $T$ is the farthest distance from its root,
and a node is at the {\em level} $\ell$ if its distance from the root
is $h-\ell$.
For a node $v$ of a rooted tree $T$, we call a {\em limb of $v$}
a subtree of $T$ rooted at some child node of $v$.
Our rooted trees are unordered,
and they ``grow top-down'', i.e.\ we depict the root on the top.
We switch from considering $m$-coloured trees to
more convenient $t$-labelled ones, the difference being that 
one vertex may have several labels at once (and so $m\sim 2^t$).
We say that two such rooted labelled trees are {\em l-isomorphic} if there is 
an isomorphism between them preserving the root and all the labels.

For obtaining the desired kernel,
we shall use a concept of reducing a (rooted $t$-labelled) tree $T$ as follows.
\begin{defi}[$f$-reduction, $f$-reduced]
Let $f:\mathbb N\to\mathbb N$ be a function, called a {\em threshold function}.
Assuming a node $v\in V(T)$ at level $i+1>0$
and a limb $B$ of $v$ such that there exist at least 
$f(i)$ other limbs of $v$ in $T$ which are all l-isomorphic
to $B$; we say that $T$ {\em$f$-reduces in one step} to $T-V(B)$.
A tree $T$ {\em$f$-reduces to} $T_0\subseteq T$ if there is a sequence
of one-step $f$-reductions from $T$ to~$T_0$.
The tree $T_0$ is {\em$f$-reduced} if no further $f$-reduction step is possible.
\end{defi}
A straightforward induction shows that any $f$-reduced $t$-labelled
rooted tree of fixed height has size bounded with respect to $f$ and~$t$.

\subsection{Reduction to kernel}
\label{sub:msokernel}

Considering \MSOtree sentences with $q$ element quantifiers and $s$ set quantifiers,
we use the following threshold function $R$ to define our reduced kernel
(where $k$ is an arbitrary integer parameter that will count all the
labels used in our proof):
\def\threshdef{%
\begin{eqnarray}
\Trec iqsk &=& q\cdot {\Nrec iqsk}^s
	,\qquad\mbox{where}
\label{eq:NR1}\\[3pt]\nonumber
\Nrec0qsk &=& 2^k+1 \geq 2  \quad\mbox{and}
\\\label{eq:NR2}
\Nrec{i+1}qsk &=& 2^k\cdot \big(\Trec iqsk +1 \big)^{\Nrec iqsk}
	\leq 2^k\cdot \big(2q\cdot {\Nrec iqsk}^s \big)^{\Nrec iqsk}\quad
\end{eqnarray}
}%
\threshdef
Note also that all these values are non-decreasing in the parameters
$q,s,k$.

The idea behind our approach can be simplified as follows.
Fix $i\geq0$ and any $a\geq\Trec iqs{k}$ where $k=t+3q+s$,
and choose an arbitrary $\Trecf qsk$-reduced rooted $t$-labelled tree $U$ 
of height $i$.
Then no \MSOtree sentence with $q$ element variables and
$s$ set variables can distinguish between $a$ disjoint copies and 
$a+1$ disjoint copies of~$U$.
Our full result then reads:

\begin{thm}
\label{thm:dtreekern}
Let $T$ be a rooted $t$-labelled tree of height $h$,
and let $\phi$ be an \MSOtree sentence with $q$ element quantifiers and
$s$ set quantifiers.
Suppose that $u\in V(T)$ is a node at level $i+1$ where $i<h$.
\begin{enumerate}[label={\bf\alph*)}]
\item If, among all the limbs of $u$ in $T$, 
there are more than $\Trec iqs{t+3q+s}$ pairwise l-isomorphic ones,
then let $T'\subseteq T$  be obtained by deleting one of the latter limbs from $T$.
Then, $T\models\phi$ $\iff$ $T'\models\phi$.

\item Consequently, the tree $T$ $\,\Trecf qs{t+3q+s}$-reduces to
$T_0\subseteq T$ such that $T_0$ itself is $\Trecf qs{t+3q+s}$-reduced, 
and $T\models\phi$ $\iff$ $T_0\models\phi$.

\item The tree $T_0$ can be computed in linear time from
$T$ and $\phi$, and the size of $T_0$ is bounded by
$$|V(T_0)|\leq\towf{h}\big[(2^{h+5}-12)\cdot (t+q+s)(q+s)\big].$$
\end{enumerate}
\end{thm}

\noindent In the latter expression, 
$\towf i(x)$ denotes the $i$-fold exponential function defined inductively as follows:
$\towf0(x)=x$ and $\towf{i+1}(x)=2^{\towf{i}(x)}$.
Then $\towf h(x)$ is an elementary function of $x$ for each
particular height~$h$.
Note that the runtime bound in c) is within classical complexity---not
parameterized.
\smallskip

In the case of FO logic, a statement analogous to Theorem~\ref{thm:dtreekern}
can be obtained using folklore arguments of finite model theory
(even full recursive expansion of all $q$ vertex quantifiers in $\phi$ could
``hit'' only bounded number of limbs of $u$ and the rest would not matter).
However, in order to obtain suitable explicit bounds, 
in the case of \MSO logic there are additional nontrivial complications
which require careful considerations (in addition to standard tools) during the proof.
Briefly saying, one has to recursively consider the internal structure
of the limbs of $u$, and show that even an expansion of a vertex-set
quantifier in $\phi$ does not effectively distinguish too many of them
(and hence some of them remain irrelevant for the decision whether
$T\models\phi$).

Before proceeding with formal proof of Theorem~\ref{thm:dtreekern}, we need
to clarify the meaning of the values $N$:

\begin{lem}
\label{lem:Nreccount}
For any natural $i,q,s$, and $k$,
there are at most $\Nrec{i}qsk$ pairwise non-l-isomorphic $\Trecf qsk$-reduced 
rooted $k$-labelled trees of \mbox{height~$\leq i$.}
\end{lem}
\proof
This claim readily follows from Equation \eqref{eq:NR1} and \eqref{eq:NR2}
by induction on~$i$.
The base case $i=0$ is trivial, and the count includes also the empty tree.
A rooted $k$-labelled tree $T$ of height $\leq i+1$
can be described by the label of its root $r$ ($2^k$~possibilities),
and the set of its limbs, each one of height $\leq i$.
This set of limbs can be fully described by the numbers of limbs
(between $0$ and $\Trec{i}qs{k}$) in every of $\leq\Nrec{i}qs{k}$
possible l-isomorphism classes.
Hence by \eqref{eq:NR2} we have got at most $\Nrec{i+1}qsk$ possible distinct
descriptions of~$T$.\qed

\smallskip
\proof[{Proof of Theorem~\ref{thm:dtreekern}}]
For clarity, we start with a proof sketch.
\begin{enumerate}[label=(\Roman*)]
\item\label{it:qelim}
We are going to use a so called ``quantifier elimination'' approach.%
\footnote{This approach has been inspired by~\cite{DKT10},
though here it is applied in a wider setting of \MSO logic.}
That means, assuming $T\models\phi$ \hbox to 0pt{$~~~\,\not$}$\iff$
$T'\models\phi$, we look at the ``distinguishing choice'' of the first
quantifier in $\phi$, and encode it in the labelling of $T$
(e.g., when $\phi\equiv\exists x.\psi$, we give new exclusive labels to
the value of $x$ and to its parent/children in $T$ and $T'$).
By an inductive assumption, we then argue that the shorter formula $\psi$
cannot distinguish between these newly labelled $T$ and $T'$, which is a
contradiction.
\item\label{it:replimb}
The traditional quantifier elimination approach---namely of set quantifiers in
$\phi$, however, might not be directly applicable to even very many 
pairwise l-isomorphic limbs in $T$ if their size is unbounded.
Roughly explaining, the problem is that a single valuation of a set variable on
these repeated limbs may potentially pairwise distinguish all of them.
Hence additional combinatorial arguments are necessary to bound the
size of the limbs in consideration.
\item
Having resolved technical (\ref{it:replimb}),
the rest of the proof is a careful composition of inductive arguments
using the formula \eqref{eq:NR1} $\Trec iqsk = q\cdot {\Nrec iqsk}^s$.
\end{enumerate}\smallskip

\noindent {\bf a)}
The whole proof goes through by {\em means of contradiction}.
That is, we assume $T\models\phi$ \mbox{while~$T'\models\neg\phi$}
(a counterexample to Theorem~\ref{thm:dtreekern}\,a, where $T'$ implicitly
depends on the choice of $u$),
up to natural symmetry between $\phi$ and $\neg\phi$ in this context.
Let $t'=t+3q+s$.
Let $B_1,\dots,B_p\subseteq T$ where $p>\Trec iqs{t'}\geq1$
be the pairwise l-isomorphic limbs of $u$ in $T$,
as anticipated in Theorem~\ref{thm:dtreekern}\,a).
Note that the height of $B_1$ is at most $i$ (but it may possibly be lower
than $i$).

So, say, $T'=T-V(B_1)$.
We will apply nested induction, primarily targeting the structure
of the sentence $\phi$, or simply the value $q+s$.
For that we assume $\phi$ in the prenex form,
i.e., with a leading section of all quantifiers.
If $q=s=0$, then $\phi$ is a propositional formula which evaluates to true
or false without respect to $T$ or $T'$.
Hence we further assume $q+s>0$.
Note also the little trick with choice of $t'=t+3q+s$ which ``makes room''
for (\ref{it:qelim}) adding further labels to $T$ in the course of the proof.

\gdef\pparagraph#1{\medskip\noindent{\it#1 }}

\pparagraph{(Minimality setup)}
To overcome the complication in (\ref{it:replimb}), 
we have to deal with limbs $B_1,\dots,B_p$ of bounded size.
So, among all the assumed counterexamples to Theorem~\ref{thm:dtreekern}\,a)
for this particular $\phi$ or symmetric $\neg\phi$, 
choose one (meaning precisely the choice of $T$ and $u$ within it)
which minimizes the size of $B_1$ (same as the sizes of $B_2,\dots,B_p$).
This minimality choice actually represents a secondary induction in our proof.

We would like to show that the l-isomorphic limbs $B_1,\dots,B_p$ are
$\Trecf qs{t'}$-reduced.
Suppose not, and let $w_k\in V(B_k)$ be a node at level $j+1\leq i$ such that
among all the limbs of $w_k$ in $B_k$
there are more than $\Trec jqs{t'}$ pairwise l-isomorphic ones,
hereafter denoted by $D_{k,1},\dots,D_{k,r}$ where $r>\Trec jqs{t'}$.
This choice is made for all $k=1,\dots,p$ symmetrically,
i.e., all the subtrees $B_k^-=B_k-V(D_{k,1})$ where $k=1,\dots,p$
are pairwise l-isomorphic, too.

We define a sequence of trees by
$U_0=T$ and $U_{k}=U_{k-1}-V(D_{k,1})$ for $k=1,\dots,p$.
Recall that $U_0\models\phi$.
If it ever happened that $U_{k-1}\models\phi$ but $U_{k}\models\neg\phi$,
then we would consider $U_{k-1}$ and $w_k$ in place of $T$ and $u$
above, and hence contradict the choice minimizing $B_1$
(which would be replaced with smaller $D_{k,1}$).
We may thus say that $U_p\models\phi$.
We similarly define $U_1'=T'$ and $U_{k}'=U_{k-1}'-V(D_{k,1})$ for
$k=2,\dots,p$ (recall that $B_1$ has been removed from~$T'$).
With an analogous argument we conclude that $U_p'\models\neg\phi$.

Note that, now, $B_1^-,\dots,B_p^-$ are pairwise l-isomorphic limbs of $u$
in $U_p$, and they are strictly smaller than $B_1$.
Since $U_p'=U_p-V(B_1^-)$, we may have chosen $U_p$ and $u$ in place of
$T,u$, again contradicting minimality of $B_1$ in the choice above.
Indeed, the (original) limbs $B_1,\dots,B_p$ are $\Trecf qs{t'}$-reduced in $T$.

\pparagraph{(Quantifier elimination: $\exists x$)}
As the main induction step
we now ``eliminate'' the leading quantifier of $\phi$ as follows.
Suppose first that $\phi\equiv\exists x.\,\psi$.
Let $a\in V(T)$ be such that $T[x=a]\models\psi(x)$.
Clearly, it can be chosen $a\not\in V(B_1)$
since $B_1$ is l-isomorphic to other $B_2,\dots,B_p$.
On the other hand, $T'[x=b]\not\models\psi(x)$ for all $b\in V(T')$.

We define a $(t+3)$-labelled tree $T^a$ which results from $T$
by adding a new label $L_x$ exclusively to the node $a$,
a new label $L_{px}$ exclusively to the parent node of $a$,
and $L_{cx}$ to the child nodes of~$a$.
A tree ${T^a}'=T^a-V(B_1)$ is formed analogously from $T'$.
Then we translate the formula $\psi(x)$ with free $x$ 
into a closed one $\psi^x$ as defined next:
All label predicates $L(x)$ in $\psi(x)$ are simply evaluated as $L(a)$ over~$T$
(which is the same as over $T'$).
Any predicate $x=y$ is replaced with $L_x(y)$.
Finally, all predicates for edges $(x,y)$ and $(y,x)$ in this parent-child order
are replaced with $L_{cx}(y)$ and $L_{px}(y)$, respectively.
It is trivial that $T[x=a]\models\psi(x)$
$\iff$ $T^a\models\psi^x$, and $T'[x=a]\not\models\psi(x)$
$\iff$ ${T^a}'\not\models\psi^x$.

All the limbs $B_1,\dots,B_p$ remain pairwise l-isomorphic in $T^a$ 
unless, say, $a\in V(B_p)$.
Even in the latter case we anyway obtain, using \eqref{eq:NR1}, at least 
$p-1>\Trec iqs{t'}-1$ pairwise l-isomorphic limbs of $u$ in $T^a$, including~$B_1$.
It is
$$\Trec iqs{t'}-1=q\cdot {\Nrec iqs{t'}}^s -1\geq
(q-1)\cdot {\Nrec iqs{t'}}^s\geq\Trec i{q-1}s{t'}
\,.$$
Note also that $q-1$ is the number of element quantifiers in $\psi$,
and that the combined parameter $t+3+3(q-1)+s=t+3q+s=t'$ remains the same.
Hence we can apply the inductive assumption to $T^a,u$, and $\psi^x$---%
concluding that $T^a\models\psi^x$ $\iff$ ${T^a}'\models\psi^x$,
a contradiction.

\pparagraph{(Quantifier elimination: $\exists X$)}
We are finally getting to the heart of the proof.
Suppose now that $\phi\equiv\exists X.\,\psi$.
Let $A\subseteq V(T)$ be such that $T[X=A]\models\psi(X)$.
On the other hand, $T'[X=A']\not\models\psi(X)$ for all $A'\subseteq V(T')$.
We define a $(t+1)$-labelled tree $T^A$ which results from $T$
by adding a new label $L_X$ precisely to all members of~$A$.
Then we translate the formula $\psi(X)$ with free $X$
into a closed one $\psi^X$ by replacing every occurrence of $y\in X$
with $L_X(y)$.
Trivially, $T[X=A]\models\psi(X)$ $\iff$ $T^A\models\psi^X$.

Note again that $s-1$ is the number of set quantifiers in $\psi$,
and that the combined parameter $t+1+3q+(s-1)=t+3q+s=t'$ remains the same.
A key observation is that ``casting'' the new label $L_X$ onto the limbs
$B_1,\dots,B_p$ may create at most $\Nrec iqs{t'}$ l-isomorphism classes
among them.
This is simply because, for each $k=1,\dots,p$,
the corresponding $B_k^A$ carries $t+1\leq t'$ labels,
it is of the same height as $B_k$ and $\Trecf qs{t'}$-reduced, too.
Hence, altogether, there are at most $\Nrec iqs{t'}$ pairwise
non-l-isomorphic choices for such $B^A_k$ by Lemma~\ref{lem:Nreccount}.

So, among all $B_1,\dots,B_p$, there are at least $p/\Nrec iqs{t'}$ pairwise
l-isomorphic limbs, and using \eqref{eq:NR1},
$$\frac p{\Nrec iqs{t'}} \>>\> \frac{\Trec iqs{t'}}{\Nrec iqs{t'}} \>=\>
 q\cdot {\Nrec iqs{t'}}^{s-1}\geq \Trec iq{s-1}{t'}
.$$
For simplicity, let the latter limbs be $B_1,\dots,B_{p'}$ where
\vspace{1pt}%
$p\geq p'>\Trec iq{s-1}{t'}$.
Now we apply the inductive assumption to $T^A,u$, and $\psi^X$.
Up to symmetry between the limbs, we get $(T^A)'=T^A-V(B_1)$ such that
$T^A\models\psi^X$ $\iff$ $(T^A)'\models\psi^X$.
Now we can define $A'\subseteq V(T')$ as the set of those nodes having label
$L_X$ in $(T^A)'$, and hence
$(T^A)'\models\psi^X$ $\iff$ $T'[X=A']\models\psi(X)$,
a contradiction to the initial assumption.

\pparagraph{(Quantifier elimination: $\forall$)}
Finally, the cases of universal quantifiers in $\phi$ are solved analogously
($\neg\exists$ in place of $\forall$).

\medskip\noindent{\bf b)}
This part readily follows by a recursive bottom-up application of a) to the whole
tree~$T$.

\medskip\noindent{\bf c)}
$T_0$ is easily constructed from $T$
by a natural adaptation of the classical linear-time tree-isomorphism algorithm.
Notice that there are no ``hidden huge constants'' depending on $\phi$ 
in this algorithm;
we simply read the parameters $t,q,s$ by linear-time parsing of $\phi$
and we compute the values of the threshold function $\Trecf qs{t+3q+s}$ 
``on demand'' when we encounter a limb at level $i$ with (too) many
l-isomorphic siblings.
We use the simple fact that $\Trec iqsk\geq\Trec{i-1}qsk$ and this computation is
thus negligible compared to the size of $T$.

Since $T_0$ is $\Trecf qs{t'}$-reduced by b), where $t'={t+3q+s}$,
we can consider $T_0\subseteq U_{h,q,s,t}$
where $U_{h,q,s,t}$ is the ``maximal'' $\Trecf qs{t'}$-reduced
rooted $t$-labelled tree of height $h$:
$U_{h,q,s,t}$ contains (at each level $j+1$) precisely $\Trec{j}qs{t'}$ limbs
of every l-isomorphism class of rooted $t'$-labelled trees of height $\leq j$.

Therefore, by Lemma~\ref{lem:Nreccount}, the number of descendants at each 
level $j$ of $U_{h,q,s,t}$ is at most $\Trec{j-1}qs{t'}\cdot\Nrec{j-1}qs{t'}$.
The total number of vertices in $U_{h,q,s,t}$ is at most
$$
1+\Trec{h-1}qs{t'}\cdot\Nrec{h-1}qs{t'}\cdot\big(
	1+\Trec{h-2}qs{t'}\cdot\Nrec{h-2}qs{t'}\cdot(1+\dots)\big)\leq
$$ \begin{equation}\label{eq:Usize}
\qquad\leq \prod_{i=0}^{h-1}\big(1+ \Trec{i}qs{t'}\cdot\Nrec{i}qs{t'}\big)
\,.\end{equation}

The task is now to estimate, by induction on $i$,
the value $1+\Trec{i}qs{t'}\cdot\Nrec{i}qs{t'}$ from above by
$\towf{i+1}\big[(6\cdot2^i-2)t'(q+s)\big]$.
Note that $t'\geq q+s\geq 1$.

\begin{eqnarray*}
1+\Nrec{0}qs{t'}\cdot\Trec{0}qs{t'} &=&
	1+q\cdot \left(2^{t'}+1\right)^{s+1}
	\leq 2^q\cdot 2^{(t'+1)(s+1)} \leq
\\
&\leq&
	2^{2t'(s+1)+q} \leq 2^{2t'(s+1+q)} \leq 2^{4t'(s+q)}
\end{eqnarray*}
\begin{eqnarray}
1+\Nrec{i+1}qs{t'}\cdot\Trec{i+1}qs{t'} &=& 1+q\cdot \Nrec{i+1}qs{t'}^{s+1} =
\label{eq:Nreccalcu}\\\nonumber
&=&	1+q\cdot\left[2^{t'}\cdot \big(\Trec iqs{t'} +1 \big)^{\Nrec iqs{t'}}\right]^{s+1}
	\leq
\\\nonumber
&\leq&	1+q\cdot\left[2^{t'}\cdot \big(2\Trec iqs{t'} \big)^{\Nrec iqs{t'}}\right]^{s+1}
	\leq
\\\nonumber
&\leq&	1+q\cdot\left[2^{t'}\cdot \big(2q{\Nrec iqs{t'}}^s \big)^{\Nrec iqs{t'}}\right]^{s+1}
	\leq
\\\nonumber
&\leq&	1+q\cdot\left[\big({\Nrec iqs{t'}}^{t'+q+s} \big)^{\Nrec iqs{t'}}\right]^{s+1}
	\leq
\\\nonumber
&\leq&	1+q\cdot\left[ 2^{(t'+q+s)\cdot{\Nrec iqs{t'}}^2}\right]^{s+1}
	\leq
\\\nonumber
&\leq&	\towf1{\left[ q+(t'+q+s)(s+1)\cdot{\Nrec iqs{t'}}^2 \right]}	\leq
\\\nonumber
&\leq&	\towf1{\left[ 2^{2(t'+q+s)}\cdot{\Nrec iqs{t'}}^2 \right]}	\leq
\\\nonumber
&\leq&	\towf1{\left[ 2^{2(t'+q+s)}\cdot2^{2 \towf i
		\left((6\cdot2^i-2)t'(q+s) \right)} \right]}	\leq
\\\nonumber
&\leq&	\towf2{\left[ 2 \towf i\left((6\cdot2^i-2)t'(q+s) \right) +2(t'+q+s)
		\right]}	\leq
\\\nonumber
&\leq&	\towf2{\left[ \towf i\left(2\cdot(6\cdot2^i-2)t'(q+s) +(t'+q+s) \right)
		\right]}	\leq
\\\nonumber
&\leq&	\towf2{\left[ \towf i\left(2\cdot(6\cdot2^i-2)t'(q+s) +2t'(q+s) \right)
		\right]}	=
\\\nonumber
&=&	\towf2{\left[ \towf i\left((6\cdot2^{i+1}-2)t'(q+s) \right)
		\right]}	=
\\\nonumber
&=&	\towf{i+2}{\left[ (6\cdot2^{i+1}-2)t'(q+s) \right]}
\end{eqnarray}
With \eqref{eq:Usize}, we then get
\begin{eqnarray*}
|V(T_0)|\leq |V(U_{h,q,s,t})|&\leq&
	\prod_{i=0}^{h-1}\big(1+ \Trec{i}qs{t'}\cdot\Nrec{i}qs{t'}\big)
\leq
\\   &\leq&
	\prod_{i=0}^{h-1} \towf{i+1}\big[(6\cdot2^i-2)t'(q+s)\big]  \leq
\\   &\leq&
	\towf{h}\big[2\cdot(6\cdot2^{h-1}-2)t'(q+s)\big]  \leq
\\   &\leq&
	\towf{h}\big[(2^{h+3}-4)\cdot (t+3q+s)(q+s)\big]  \leq
\\   &\leq&
	\towf{h}\big[(2^{h+5}-12)\cdot (t+q+s)(q+s)\big]\rlap{\hbox to
          65 pt{\hfill\qEd}}
\,.\end{eqnarray*}

\begin{cor}
\label{cor:dtreealg}
Let $T$ be a rooted $t$-labelled tree of constant height $h\geq1$,
and let $\phi$ be an \MSOtree sentence with $r$ quantifiers.
Then $T\models\phi$ can be decided by an fpt algorithm running in elementary time
$$
\ca O\left(  \towf{h+1}\big[2^{h+5}\cdot r(t+r)\big]  + |V(T)|\right)
= \towf{h+1}\big(\ca O(|\phi|)^2\big) + \ca O\left(|V(T)|\right)
.$$
\end{cor}

\proof
Let $T_0$ be the kernel obtained in linear time by Theorem~\ref{thm:dtreekern}.
We (by brute force) exhaustively expand all the quantifiers of $\phi$
into all possible valuations in $T_0$, 
having at most $2^{|V(T_0)|}$ possibilities for each.
By searching this ``full valuation tree'' in time 
$\ca O\big(2^{|V(T_0)|\cdot(r+1)}\big)$ we decide whether $T_0\models\phi$.
Using the size bound on $T_0$ given by Theorem~\ref{thm:dtreekern}, 
where $r=q+s$, it is
\begin{eqnarray*}
2^{|V(T_0)|\cdot(q+s+1)} &\leq&
	2^{\towf{h}\!\big[(2^{h+5}-12)\cdot (t+q+s)(q+s)\big]\cdot(q+s+1)}
 \leq
\\   &\leq&
	\towf{h+1}\big[(2^{h+5}-12)\cdot (t+q+s)(q+s)+(q+s+1)\big]
\\   &\leq&
	\towf{h+1}\big[2^{h+5}\cdot r(t+r)\big]\rlap{\hbox to 182 pt{\hfill\qEd}}
\,.
\end{eqnarray*}

\subsection{Counting \MSO logic}
\label{sub:countingkernel}

Theorem~\ref{thm:dtreekern} can be further strengthened by considering
\CMSOtree logic.
Although relatively easy, this is not a simple corollary, and the additional
issues of different sort require us to repeat the overall structure 
of the previous proof as follows.

\begin{thm}
\label{thm:dtreekerncounting}
Let $T$ be a rooted $t$-labelled tree of height $h$,
let $\phi$ be a \CMSOtree sentence with $q$ element quantifiers and
$s$ set quantifiers, and let $M$ be the least common multiple of the $b$
values of all $\modp ab$ predicates occurring in $\phi$.

a) Suppose that $u\in V(T)$ is a node at level $i+1$ where $i<h$.
If, among all the limbs of $u$ in $T$, 
there are more than $\Trec i{M+q}s{t+3q+s}$ pairwise l-isomorphic ones,
then let $T'\subseteq T$  be obtained by deleting exactly $M$ of the latter limbs
from~$T$.
Then, $T\models\phi$ $\iff$ $T'\models\phi$.

b) Consequently, there is a subtree $T_0\subseteq T$ computable in linear
time (non-parameterized), such that
$T\models\phi$ $\iff$ $T_0\models\phi$ and the size of $T_0$ is bounded by
$$|V(T\smallskip_0)|\leq\towf{h}\big[(2^{h+5}-12)\cdot (t+M+q+s)(M+q+s)\big]
.$$
\end{thm}

\begin{proof}
We closely follow the structure of the proof of Theorem~\ref{thm:dtreekern},
and implicitly refer to its assumptions and notation.
In particular, let $t'=t+3q+s$.
Due to the effects of (future) quantifier elimination onto the $\modp ab(X)$
predicates we have to deal in $\phi$ also with special model constants 
$\modp ab^L$ (where $L$ is a label of the model):
The semantics of $\modp ab^L$ in $U\models\psi$ is that the respective model
$U$ contains $a$ modulo $b$ nodes holding label~$L$.

{a)}
As in the previous proof, we have a node $u$ in $T$ such that
$u$ has many, $p>\Trec i{M+q}s{t'}\geq M$, 
pairwise l-isomorphic limbs $B_1,\dots,B_p\subseteq T$.
We claim that for $T'=T-V(B_1\cup\dots\cup B_M)$ it holds
$T\models\phi$ $\iff$ $T'\models\phi$.
This is again proved by induction on $q+s$:

For the base of induction, when $q=s=0$,\, $\phi$ is a propositional formula
and the outcomes of $T\models\phi$ and $T'\models\phi$ might differ only in
the constants $\modp ab^L$.
However, $M$ is a multiple of $b$ by definition 
and the number of (deleted) nodes holding
label $L$ in $B_1\cup\dots\cup B_M$ is a multiple of $M$,
and hence all the involved model constants $\modp ab^L$ in $\phi$
indeed do have the same value over $T$ as over~$T'$.

For the induction step with $q+s\geq1$,
we show that the threshold value $p>\TrecM iqs{t'}:= \Trec i{M+q}s{t'}$ is
sufficient in this proof.
Note that $\TrecM i00{t'}=M$.
We again proceed by means of contradiction;
assuming $T\models\phi$ while $T'\models\neg\phi$ up to symmetry between
$\phi$ and $\neg\phi$.
Let us consider a counterexample which minimizes the size of $B_1$.
Then, analogously to the proof of Theorem~\ref{thm:dtreekern},
the l-isomorphic limbs $B_1,\dots,B_p$ are $\TrecfM qs{t'}$-reduced
or a smaller counterexample exists (for this $\phi$ or $\neg\phi$).
The only difference in the argument is that we are now always removing 
$M$-tuples of l-isomorphic limbs instead of single ones.

Then a quantifier elimination argument, essentially same as previous,
finishes the induction step.
The argument from the proof of Theorem~\ref{thm:dtreekern} can be simply repeated word by word, 
only replacing $q$ with $q':=M+q$ as in $\Trec i{q'}s{t'}$.

\smallskip b)
By a recursive bottom-up application of a) to the whole tree~$T$
we obtain a tree $T_0$ which is $\Trecf {q'}s{t'}$-reduced,
and this $T_0$ is computable in linear time, too.
The size bound then follows from Theorem~\ref{thm:dtreekern}c) for 
$q'=M+q$ in place of~$q$.
\end{proof}

\begin{cor}
\label{cor:dtreealgcounting}
Let $T$ be a rooted $t$-labelled tree of constant height $h\geq1$,
and let $\phi$ be an \CMSOtree sentence.
Let $M$ be the least common multiple of the $b$
values of all $\modp ab$ predicates occurring in $\phi$.
Then $T\models\phi$ can be decided by an fpt algorithm running in elementary time
$$
\towf{h+1}\big(\ca O(M+|\phi|)^2\big) + \ca O\left(|V(T)|\right).\eqno{\qEd}
$$
\end{cor}

\section{Algorithmic Consequences for Graphs}
\label{sec:algoEMSO}

If one considers extending algorithmic scope of the previous section
to richer classes of structures such as general graphs,
the first natural choice would be to employ efficient interpretation of
the structures in coloured trees of fixed height together with
Corollaries~\ref{cor:dtreealg} and~\ref{cor:dtreealgcounting}.
This has been the course taken in the conference version of this paper
(covering Lampis~\cite{Lam12} and Ganian~\cite{Gan11} as special cases) and
in a greater generality in~\cite{GHNOOR12}.
Here we use another, slightly more complicated, approach which has the
advantage of being able to smoothly incorporate also some wider problem frameworks,
such as \MSO enumeration and the \LinEMSO optimization framework,
and others.

\subsection{Smaller tree automata for \MSO}

Our approach uses tree automata for \MSO properties (cf.~Rabin~\cite{Rab69}, Doner~\cite{Doner} and Thatcher and Wright~\cite{Thatcher_Wright}), 
and its core idea is to give a stricter bound on the number of
states of such an automaton by showing that each reachable state is
represented by some of our reduced kernels.
Recall that the number of states of the automaton related to, e.g., Courcelle's
\MSOii theorem \cite{cou90} grows non-elementarily with the quantifier alternation
depth of the formula, and that this is generally unavoidable by
Frick and Grohe \cite{FG04} already for \MSO properties on trees.

However, consider the following situation;
we apply the algorithm of, say, Courcelle's \MSOii theorem to a class $\cf C$
of graphs which, in an addition to having bounded tree-width, has an
interpretation $J$ in a class of trees of bounded height.
Although the related automaton $\ca A$ has a non-elementary number of states in
general, our Theorem~\ref{thm:MSOautosmall} will show that the number of
states of $\ca A$ reachable by the graphs from $\cf C$ is indeed elementary
(in the input formula).
The important part of the formulation of Theorem~\ref{thm:MSOautosmall} is
that we can blindly bound the number of reachable states of $\ca A$,
without knowing or changing the kind of ``tree-structured'' decomposition 
(cf.\ the arbitrary interpretation $I$ versus our $J$ in the statement of
Theorem~\ref{thm:MSOautosmall})
used in the original algorithmic metatheorems.

For a class of relational structures $\cf S$,
we say that an \MSO interpretation $I$ of $\cf S$ in a class of rooted trees 
$\cf T$ is {\em hereditary} if, roughly saying,
subtrees interpret respective induced substructures.
Formally, for any $S\in \cf S$ which is interpreted in $T\in\cf T$ with the
domain $D\subseteq V(T)$, every rooted subtree $T'\subseteq T$ interprets by $I$
a structure $S'$ that is the restriction of $S$ onto $D\cap V(T')$.
Our core claim is now formulated as follows.

\begin{thm}
\label{thm:MSOautosmall}
Let $\cf S$ be a hereditary class of binary relational structures, $\psi$ a \CMSO property
over $\cf S$, and $I$ a hereditary \CMSO interpretation of $\cf S$ in a class
$\cf T$ of rooted (labelled) binary trees.
If there exists a hereditary \CMSO interpretation $J$ of $\cf S$ in the class 
$\cf U_d$ of rooted labelled trees of fixed height $d$, then the following holds:
There is a finite deterministic tree automaton $\ca A_{\psi,I}$ accepting
the language of those $T\in\cf T$ such that $T\models\psi^I$
($\psi$~under the interpretation $I$);
and the number of states of $\ca A_{\psi,I}$ is at most
\begin{equation}\label{eq:autoelementary}
\towf{d+1}{\left( \ca O(M+|\psi^J|+\iota)^2\, \right)}
,\end{equation}
where (i) $\psi^J$ is $\psi$~under the interpretation $J$,
(ii) $M$ is the least common multiple of the $b$
values of all $\modp ab$ predicates occurring in~$\psi^J$,
and (iii) $\iota$ is a constant depending on $I,\cf T$ but not on $\psi$.
\end{thm}

Recall, again, that sole existence of the automaton $\ca A_{\psi,I}$ follows from
Rabin's~\cite{Rab69}, while the selling point of this claim is an elementary
size bound on~$\ca A_{\psi,I}$ for each fixed~$d$.
Note also that we intentionally formulate the theorem with {\em two}
interpretations $I$ and $J$ which are not mutually related in any way
(even though an existence of $I$ can be easily derived from the existence
of~$J$).
The purpose of this ``separation of $I$ from $J$'' 
has been explained at the beginning of this section.

\begin{proof}
Our approach builds upon the classical 
Myhill--Nerode regularity tool in automata theory. 
We actually apply its tree-automata version~\cite{myhill-nerode} -- we show
that the number of classes of the congruence relation on trees defined with
respect to $\psi^{I}$ is bounded as in the statement of the theorem.  The
existence of the automaton $\ca A_{\psi,I}$ with the same number of states
then follows.

For a \CMSO sentence $\phi$ we define an
equivalence relation $\sim_\phi$ on the universe of rooted (labelled) trees as follows.
If $T$ and $B_2$ are rooted trees, and $B_1$ is a limb of a node $v\in
V(T)$, then let $T[B_1\to B_2]$ denote the tree obtained from $T-V(B_1)$ by
attaching the root of $B_2$ as a child of $v$.
For this proof, we additionally treat also the case of an ``improper limb''
$B_1=T$ and then, specially, $T[B_1\to B_2]=B_2$.
It is
$$ B_1 \sim_\phi B_2 \mbox{\qquad if, and only if,\qquad}
	T\models\phi\iff T[B_1\to B_2]\models\phi
$$
over all rooted trees $T$ such that $B_1$ is a limb of $T$ or $B_1=T$.

We are interested in the equivalence classes of $\sim_{\psi^I}$ when
restricted to rooted subtrees of the members of~$\cf T$.
Let $S\in\cf S$ be interpreted by $I$ in a tree $T\in\cf T$
(i.e., $S\simeq T^I$), and let $B$ be a limb in~$T$.
Let $S_1$ and $S_2$ be the induced substructures of $S$ 
interpreted in $B$ and $T-V(B)$, respectively.
Let $U\in\cf U_d$ be an interpretation of $S\simeq U^J$ under $J$.
Let $U_1,U_2$ be two disjoint copies of $U$, and let
$U_3$ be the rooted tree (of height $d+1$) obtained from $U_1\cup U_2$ 
by adding a new root as the parent of the former roots of $U_1,U_2$.

\smallskip
Our first claim is that there exists an assignment of additional labels
to $U_3$ and an interpretation $J_3$ of $S$ in $U_3$ such that;
(a) $J_3$ depends only on $J$ and on $I,\cf T$ (this will define our $\iota$),
(b) $S_i\simeq U_i^{J_3}$ for $i=1,2$,
and (c) the additional labelling of $U_1$ within $U_3$ is independent of $T-V(B)$.

To prove this claim, consider any of the binary relational symbols $R$ over $\cf S$
and the \CMSO formula $\varrho(x,y)\equiv R(x,y)^I$ which interprets $R$
into $\cf T$.
Already by Rabin's theorem~\cite{Rab69}---while simply amended by finite-state
$\modp ab$ predicates in $\varrho$\,---%
this $\varrho(x,y)$ has a finite number of equivalence classes
with the interpretation of $x$ in $V(B)\cap dom(S)$ and $y$ into
$(V(T)\cap dom(S))\sem V(B)$ (and the same applies also to
$\varrho(y,x)$ interpreting the inverse $R^{-1}$ in case of a non-symmetric relation).
Hence there exists an integer $m_R$ depending only on $I,\cf T$ such that
$V(S_1)$ has a partition $\cf P_R$ with $|\cf P_R|\leq m_R$ parts, and
$S\models R(x_1,y) \iff S\models R(x_2,y)$ for any two $x_1,x_2$ from the
same part of $\cf P_R$ and any $y\in V(S_2)$.
Importantly, the partition $\cf P_R$ does not depend at all on $S_2$ 
and $T-V(B)$.

We choose $\iota=2+\sum_{R\rm\>over\>\cf S}(m_R+m_{R^{-1}})$ and assign $\iota$ new
labels to the nodes of $U_3$ as follows:
For $i=1,2$, a separate new label $L_i$ is issued to the whole domain of $S_i$ in $U_i$.
Furthermore, on the domain of $S_1$ in $U_1$, every vertex gets one of $m_R$ new labels
identifying which part of $\cf P_R$ it belongs to.
This is repeated for all relational symbols $R$ and their inverses $R^{-1}$ 
over $\cf S$.
On the domain of $S_2$ in $U_2$, the label identifying a part $P\in\cf P_R$
is simply given to all the vertices which are adjacent to $P$ via $R$.
The interpretation $J_3$ now follows naturally;
$U_3\models R(x,y)^{J_3}$ if, and only if,
\begin{itemize}
\item
 $x,y$ in the domain of $S$ are from the same one of $U_1,U_2$\,---formally, 
$L_1(x)\wedge L_1(y)$ or $L_2(x)\wedge L_2(y)$\,---%
and $U_3\models R(x,y)^{J}$ (since $J$ is hereditary), or
\item
 $x,y$ in the domain of $S$ are from different ones of $U_1,U_2$\,---formally, 
$L_1(x)\wedge L_2(y)$ or vice versa---%
and the additional labels in $U_3$ ``encode'' $R(x,y)$ as defined previously.
\end{itemize}

\noindent This finishes the proof of the first claim; the existence of $J_3$.
Note, moreover, that $T\models\psi^I\iff S\models\psi\iff
U_3\models\psi^{J_3}$.

\smallskip
Second, we claim that every equivalence class of $\sim_{\psi^I}$ contains a
special small representative (not necessarily unique),
and so the index of $\sim_{\psi^I}$ cannot be too large.
We apply Theorem~\ref{thm:dtreekerncounting}\,a) to $\psi^{J_3}$ and
the tree $U_3$, precisely to all nodes of $U_1$ within it.
Let $U_1$ reduce to $U_0$ by this application(s), and let $U_{30}=U_3[U_1\to U_0]$.
Then $U_3\models\psi^{J_3}\iff U_{30}\models\psi^{J_3}$.
Since $J$ is hereditary, $U_0^{\>J_3}$ as a restriction of $U_0^{\>J}$ 
is isomorphic to an induced substructure $S_0\subseteq S_1$.
Let $B_0\subseteq B$ be the subtree giving $dom(S_0)$ in the
interpretation~$I$, and let
$S_{30}=S-(V(S_1)\sem V(S_0))\simeq U_{30}^{\>J_3}$.
Then, again, $B_0^{\>I}\simeq S_0$ and $T[B\to B_0]^I\simeq S_{30}$
since $I$ is hereditary.
This $B_0$ will be a representative of the $\sim_{\psi^I}\,$-class of~$B$.

By the previous, we have got $T\models\psi^I\iff U_{30}\models\psi^{J_3}
  \iff S_{30}\models\psi\iff T[B\to B_0]\models\psi^I$.
This has been so far verified for one particular tree $T\in\cf T$ having $B$
as its limb.
Consider now arbitrary $T'\in\cf T$ having $B\subseteq T'$ as its limb or $B=T'$,
and correspondingly define $S'\in\cf S$, $S'\simeq {T'}^I$
and $U'\in\cf U_d$ such that $S'\simeq {U'}^J$ by the assumptions.
If $S_1',S_2'\subseteq S'$ are the induced substructures interpreted in $B$
and $T'-V(B)$, then $S_1'=S_1$.
We take $U_1':=U_1$ and $U_2'$ a disjoint copy of $U'$,
and analogously construct $U_3'$ from $U_1'\cup U_2'$ by adding a new root.
As in the first claim, and emphasizing the condition (c),
we get a labelling of $U_3'$ such that $S'\simeq U_3'^{\>J_3}$
and $S_i'\simeq{U_i'}^{J_3}$ for $i=1,2$.
Again by Theorem~\ref{thm:dtreekerncounting}\,a), 
for $U_{30}'=U_3'[U_1'\to U_0]$ it holds
$T'\models\psi^I\iff S'\models\psi
  \iff U_{3}'\models\psi^{J_3} \iff U_{30}'\models\psi^{J_3}
  \iff T'[B\to B_0]\models\psi^I$.
Consequently, $B\sim_{\psi^I}B_0$ as desired.

\smallskip
Therefore, the number of equivalence classes of $\sim_{\psi^I}$ is at most
as large as the number of pairwise non-l-isomorphic 
$\Trecf {M+q}s{t+3q+s}$-reduced rooted $k$-labelled trees 
($B_0$) of height~$\leq d$, where
$q,s$ are the numbers of element and set quantifiers in $\psi^{J_3}$
and $t$ is the number of labels addressed in $\psi^{J_3}$.
This quantity is at most $\Nrec d{M+q}s{t+3q+s}$ by Lemma~\ref{lem:Nreccount}.
Under a very rough estimate, $q,s\leq t+3q+s\leq\ca O(|\psi^J|+\iota)$
by the construction of $J_3$.
Using the calculation of \eqref{eq:Nreccalcu} (for $i+1=d$, and again with 
a broad margin) we get that $\Nrec d{M+q}s{t+3q+s}<
  \towf{d+1}{\left( \ca O(M+|\psi^J|+\iota)^2\, \right)}$.
This is an upper bound on the number of states of desired minimal 
$\ca A_{\psi,I}$ by the Myhill--Nerode theorem.
\end{proof}

\subsection{Solving extended \MSO properties}
\label{sub:fasteralg}

Unfortunately, direct algorithmic applicability of 
Corollaries~\ref{cor:dtreealg} and~\ref{cor:dtreealgcounting}
is limited to pure decision problems (such as, e.g.,~$3$-colourability),
but many practical problems are formulated as optimization ones.
The usual way of transforming optimization problems into decision ones does
not work for us since the $\MSO$ language cannot handle arbitrary numbers.

Nevertheless, there is a known solution.
Arnborg, Lagergren, and Seese~\cite{als91} (while studying graphs of
bounded tree-width), and later Courcelle, Makowsky, and Rotics~\cite{cmr00}
(for graphs of bounded clique-width), specifically extended the expressive
power of MSO logic to define so-called \LinEMSO optimization problems.
Briefly saying, the \LinEMSO language allows, in addition to ordinary
\MSO expressions, to compare between and optimize over linear evaluational terms.

We follow, for an illustration, a simpler definition
of $\LinEMSO_1$ given in~\cite{cmr00}.
Consider any $\MSOi$ formula $\psi(X_1,\dots,X_p)$ with free set variables,
and state the following problem on an input graph $G$:
\[\label{eq:maxpsi}
\prebox{opt}\big\{ f_{lin}(Z_1,\dots,Z_p):\>
	Z_1,\dots,Z_p\subseteq V(G),~
	G\models\psi(Z_1,\dots,Z_p) \big\}
\,,\]
where $\prebox{opt}$ can be min or max, and $f_{lin}$ is a linear
evaluational function.
It is
\[\label{eq:lineval}
f_{lin}(Z_1,\dots,Z_p) ~=~
	\sum_{i=1}^{p}\sum_{j=1}^{m} \left( a_{i,j}\cdot
		\sum_{x\in Z_i}f_{j}(x)\right)
\]
where $m$ and $a_{i,j}$ are (integer) constants and $f_j$ are (integer) weight
functions on the vertices of~$G$.
Typically $f_{lin}$ is just a cardinality function.
Such as, 
\[ \psi(X)\,\equiv\, \forall v,w\, \big(
	v\not\in X\vee w\not\in X\vee \neg\prebox{edge}(v,w)\big)
	\mbox{~~~and~~~``} \max\,|X|\, " \]
describes the maximum independent set problem, or
\[\label{eq:minds}
 \psi(X) ~\equiv~ \forall v \exists w\>
	\big[v\in X \vee \big(w\in X\wedge \prebox{edge}(v,w) \big)\big]
	\mbox{~~~and~~~``} \min\,|X|\, "
\]
is the minimum dominating set problem.

The algorithms given in \cite{als91,cmr00} for solving such \LinEMSO
optimization (and enumeration as well) problems are implicitly based on a
finite tree automaton associated with the formula $\psi$ in the problem
description.
Now, Theorem~\ref{thm:MSOautosmall} can immediately be used to tighten 
runtime analysis of each of the mentioned algorithms, when the input is
restricted to graph classes having not only bounded tree-width or
clique-width, respectively, but at the same time being interpretable in a
class of trees of fixed height.

This goal first gets us to the following definition:

\begin{defi}[Tree-depth \cite{NO06}]
\label{df:tree-depth}
The {\em closure} $cl(F)$ of a rooted forest $F$ is the graph obtained from
$F$ by adding from each node all edges to its descendants.
The {\em tree-depth} $\td(G)$ of a graph $G$ is one more than the smallest height
(distance from the root to all leaves) of a rooted forest $F$ such that 
$G\subseteq cl(F)$.
\end{defi}

\begin{figure}[tb]
\center
\begin{tikzpicture}[xscale=1.4,yscale=0.8]
\tikzstyle{every node}=[draw, shape=circle, minimum size=4pt,inner sep=0pt, fill=black]
    \draw (0,6) node (a) {}
        -- ++(230:2.5cm) [color=red] node (b) {}
        -- ++(230:1.0cm) node (c) {}
        -- ++(230:0.5cm) node (d) {};
    \draw  (a) {}
        -- ++(310:2.5cm) [color=red] node (e) {}
        -- ++(230:1.0cm) node (f) {}
        -- ++(230:0.5cm) node (g) {};
    \draw  (b) {}
        -- ++(310:1.0cm) [color=red] node (h) {}
        -- ++(230:0.5cm) node (i) {};
    \draw  (e) {}
        -- ++(310:1.0cm) [color=red] node (j) {}
        -- ++(230:0.5cm) node (k) {};
    \draw  (c) {}
        -- ++(310:0.5cm) [color=red] node (l) {} ;
    \draw  (f) {}
        -- ++(310:0.5cm) [color=red] node (m) {} ;
    \draw  (h) {}
        -- ++(310:0.5cm) [color=red] node (n) {} ;
    \draw  (j) {}
        -- ++(310:0.5cm) [color=red] node (o) {} ;
    \draw (d) -- (c) -- (l) -- (b) -- (i) -- (h) -- (n) -- (a) -- (g)
	-- (f) -- (m) -- (e) -- (k) -- (j) -- (o) [thick];
\end{tikzpicture}
\caption{%
The path of length $14$ has tree-depth $3+1=4$ since it is contained in the
closure of the depicted (red) tree of height $3$.
It can be proved that this is optimal.
}
\label{fig:td-path14}
\end{figure}

Note that tree-depth is always an upper bound for tree-width.
Some useful properties of it can be derived from the following asymptotic
characterization:
If $L$ is the length of a longest path in a graph $G$, then
  $\lceil \log_2 (L+2)\rceil\leq {\rm td}(G)\leq L+1$.
See Figure~\ref{fig:td-path14}.
For a simple proof of this, as well as for a more extensive study of tree-depth, 
we refer the reader to \cite[Chapter~6]{Sparsity}. 
Here we need the following:

\begin{lem}
\label{lem:tdinterpret}
Let $d$ be an integer and $\cf R_d$ denote the class of all rooted
$d$-labelled trees of height~$d$.
The class of all graphs of tree-depth at most $d$ has
a hereditary \MSOi interpretation into $\cf R_{d+1}$.
\end{lem}
\begin{proof}
Let $\td(G)\leq d$ and $W$ be a rooted forest of height $d$ such that
$G\subseteq cl(W)$, and let $T\in\cf R_{d+1}$ be obtained from $W$ 
by adding a new common root.
The intended interpretation identically maps $V(G)$ into $V(T)$.
In particular, each vertex quantifier $\exists x\dots$ is simply
replaced with $\exists x.\neg L_0(x)\wedge\dots$,
where $L_0$ is a special label given to the root of $T$.

Every vertex $v$ of $G$ at distance $i\geq1$ from the root of $T$
is given the label~$L_i$.
Every vertex $v$ of $G$, such that there exists
an ancestor $u$ of $v$ in $T$ and $uv\in E(G)$,
is also given the label $L_j$ where $1\leq j<i$ and $j$ 
is the distance of $u$ from the root in $T$.
Note that $L_i(x)\wedge\bigwedge_{j>i}\neg L_j(x)$ is true iff $x$
is at the distance $i$ from the root.
It is a routine to express the edge relation $\eta$ of $G$ as follows:
$\eta(x,y)\>\equiv\>x\not=y\wedge(\alpha(x,y)\vee\alpha(y,x))$ where
$$\alpha(x,y)\>\equiv\>\text{ancest}(x,y)\wedge\bigvee_{i=1,\dots,d}\big[
	L_i(x)\wedge L_i(y)\wedge\bigwedge_{j>i}\neg L_j(x)\big].$$
Here $\text{ancest}(x,y)$ means that $x$ is an ancestor of $y$
(the transitive closure of the parental relation in~$T$).
This is clearly
\MSOi-expressible (in fact, even \FO-expressible on trees of bounded height).
\end{proof}

\begin{cor}
\label{cor:linemso2}
Let $\cf C$ be a class of graphs of bounded tree-depth.
Then every $\LinEMSO_2$ problem $\ca P$ can be solved on $\cf C$ by a linear-time fpt
algorithm with an elementary runtime dependence on $\ca P$.
Then same holds also if \CMSOii is allowed in the description of $\ca P$.
\end{cor}
\begin{proof}
Let the definition of $\ca P$ be based on an \CMSO formula $\psi$
(for simplicity, $\psi$ can be thought of an \CMSOi formula thanks to
Theorem~\ref{thm:MS1=MS2}).
Then, by Lemma~\ref{lem:tdinterpret}, Theorem~\ref{thm:MSOautosmall} applies
here and the runtime analysis of \cite{als91} can be tightened to an
elementary bound \eqref{eq:autoelementary} of Theorem~\ref{thm:MSOautosmall}.
\end{proof}

In exactly the same way we can claim an analogous statement for
$\LinEMSO_1$:
\begin{cor}
\label{cor:linemso1}
Let $\cf C$ be a class of graphs of bounded clique-width,
and assume that $\cf C$ has a hereditary \MSOi interpretation in
a class of trees of fixed height.
Then every $\LinEMSO_1$ problem $\ca P$ can be solved on $\cf C$ by a linear-time fpt
algorithm with an elementary runtime dependence on $\ca P$,
provided a clique-width expression for the input graph is given.
Then same holds also if \CMSOi is allowed in the description of $\ca P$.
\qed
\end{cor}

In relation to Corollary~\ref{cor:linemso1} it becomes interesting to ask what graph
classes have an \MSOi interpretation in a class of trees of fixed height.
The answer \cite{GHNOOR12} is mentioned in another context later in
Section~\ref{sec:treemodels} (Definition~\ref{def:shrub-depth}).

\section{Consequences for Expressive Power of \FO}
\label{sec:exprpower}

A non-algorithmic straightforward corollary of Theorem~\ref{thm:dtreekern}
is the fact that \FO and \MSO logic can express the same collection of properties
on classes of trees of bounded height; they have equal {\em expressive power}.
Formally:

\begin{prop}
\label{prop:samepower}
Let $h,t$ be integers.
If $\phi$ is an \MSO sentence,
then there is an \FO sentence $\psi_{h,t}$ such that, for any
rooted $t$-labelled tree $T$ of height at most~$h$, it is
$T\models\phi$ $\iff$ $T\models\psi_{h,t}$.
\end{prop}

\begin{proof}
Let $\phi$ be an \MSO sentence with $q$ element quantifiers and
$s$ set quantifiers.
By Theorem~\ref{thm:dtreekern}, there is a finite set $\ca U_\phi$ of
pairwise non-l-isomorphic \mbox{$\Trecf qs{t+3q+s}$-reduced} trees
$W$ such that $W\models\phi$, and
$T\models\phi$ if and only if the $\Trecf qs{t+3q+s}$-reduction of 
$T$ is l-isomorphic to a member of $\ca U_\phi$.

We write an \FO sentence $\psi_{h,t}\equiv \exists x.\> root(x)\wedge
	\bigvee_{W\in\ca U_\phi}\tau_W(x)$.
The intended meaning of $\tau_W$ is that $T\models\tau_W(r)$
where $r\in V(T)$ if, and only if, 
the subtree $T_r\subseteq T$ induced on $r$ and all of its descendants
reduces, up to l-isomorphism, to $W$.
Assuming existence of $\tau_W$ for a moment, we see that
$T\models\phi$ $\iff$ $T\models\psi_{h,t}$.

We build $\tau_W$ recursively by induction on the height of $W$.
For height zero, i.e.\ when $W$ is a single vertex,
$\tau_W(x)$ simply tests the correct label of $x$ and that $x$ has no
children.
Now let $W$ be of height $h>0$, with the root $w$ and its limbs
$W_{i,j}$ where $i=1,\dots,a$ and $j=1,\dots,b_i$,
such that all $W_{i,j}$ for $j=1,\dots,b_i$ are \mbox{l-isomorphic} to the
same $U_i$, and $U_i$ for $i=1,\dots,a$ are pairwise non-l-isomorphic.
Let $\ca S$ denote the set of those $U_i$ for which 
$b_i=\Trec{h-1}qs{t+3q+s}$ (the threshold in Theorem~\ref{thm:dtreekern}).

To conclude the proof, we set
\begin{eqnarray*}
\tau_W(x)&\equiv&\> \lower.5ex\hbox{\Large$\exists$}\>
	(y_{i,j}: i=1,\dots,a,\>j=1,\dots,b_i)~ \left[\,
	\bigwedge\nolimits_{i,j} \prebox{parent}(x,y_{i,j}) \wedge
	\right.
\\
	&&\wedge \bigwedge\nolimits_{i,j,i',j'} y_{i,j}\not=y_{i',j'} \wedge
	\bigwedge\nolimits_{i,j} \tau_{U_i}(y_{i,j}) \wedge
\\
	&&\wedge \left.\left( \forall z.\> \prebox{parent}(x,z) \to
	\left( \bigvee\nolimits_{i,j} z=y_{i,j} \vee
	\bigvee\nolimits_{U_i\in\ca S} \tau_{U_i}(z)
	\right)\right)\right]
\,,\end{eqnarray*}
meaning that; (1) among the limbs of $x$ in $T$ there exist pairwise
distinct ones such that, when recursively reduced, they are
in a one-to-one l-isomorphism correspondence to the limbs of $w$ in $W$;
and (2) all the other limbs of $x$ in $T$ reduce to ones 
l-isomorphic to some $U_i$ reaching the reduction threshold above.
\end{proof}

The purpose of this section is to investigate generalizations of
Proposition~\ref{prop:samepower} to richer graph classes.

\subsection{Case of bounded tree-depth}

Elberfeld, Grohe, and Tantau
\cite{EGT12} proved that \FO and \MSOii have equal expressive power on
the graphs of bounded tree-depth---Theorem~\ref{thm:EGT}.
Having Theorem~\ref{thm:dtreekern} at hand, we can provide a relatively
simple alternative proof of this result along the construction from
Proposition~\ref{prop:samepower}.
Though, in this section we take a different route, which might look
unnecessarily complicated at the first sight, but which allows for a smooth
extension to a new result about the expressive power of \FO and \MSOi
in Section~\ref{sub:fomsoshrub}.
Along this route we introduce a sophisticated combinatorial tool, 
namely well-quasi-ordering, in the logical context.

\begin{thm}[Elberfeld, Grohe, and Tantau \cite{EGT12}]
\label{thm:EGT}
Let $\cf D$ denote a class of graphs of bounded tree-depth
(Definition~\ref{df:tree-depth}).
Then \FO and \MSOii have the same expressive power on $\cf D$.
\end{thm}

\begin{proof}
Let $\psi_2$ be an \MSOii sentence.
Since \MSOi has the same expressive power as \MSOii on graphs of bounded tree-width 
(by Theorem~\ref{thm:MS1=MS2}), we may as well consider an equivalent \MSOi sentence~$\psi$.
Our alternative proof can be then outlined in three steps
(the first two of which are analogous to Proposition~\ref{prop:samepower}):
\begin{enumerate}[label=(\Roman*)]
\item
By Lemma~\ref{lem:tdinterpret}, there is a hereditary 
\MSOi interpretation $J$ of $\cf D$ in the class $\cf U=\cf U_d$ 
of rooted labelled trees of height at most $d$, for some integer constant~$d$.
By Theorem~\ref{thm:dtreekern}, there is a finite set
$\cf U_0\subseteq \cf U$ of ($f$-reduced) kernels;
every $T\in \cf U$ $f$-reduces to an ``easily definable'' $T_0\in\cf U_0$
such that $T\models\psi^J\iff T_0\models\psi^J$
(where the reduction threshold $f$ depends only on $\psi,J,d$
and we will simply say ``{\em reduces}'' in this proof).
\item
For $G\in\cf D$, hence, $G\models\psi$ is equivalent to saying
that $G\simeq T_G^{\>J}$ for a tree $T_G\in\cf U$ such that
$T_G$ reduces to a tree in
 $\cf U_0^{\psi,J}=\{\, U\in \cf U_0: U\models\psi^J \}$ 
(a~disjunction of finitely many cases over the members of $\cf U_0^{\psi,J}$).
\item
A problem is that $T_G$ is only implicit and we cannot directly
address $T_G$ (and reducibility of it to $U\in\cf U_0^{\psi,J}$) 
from within the resulting \FO formula over~$G$.
To resolve this problem, we consider a related hereditary property
over $G$ which is characterized by finitely many obstacles by 
Theorem~\ref{thm:Ding}.
We use it to build the desired \FO sentence expressing over $G$ that
(some) implicit $T_G$ reduces to a particular reduced tree $U\in\cf U_0^{\psi,J}$.
\end{enumerate}
The rest of the proof will give the details of crucial step (III).
\smallskip

Let $T\in\cf U$, and let $\ell(T)$ denote the domain (vertex set) of $T^J$.
We say that a graph $H\in\cf D$ is {\em$T$-coloured} (with respect to implicit $J$)
if $H$ is associated with an injective mapping from $\ell(T)$ into $V(H)$.
When dealing with such a coloured graph $H$ we automatically consider a subgraph
relation preserving these colours.
The key definition is now that of consistency with a $T$-colouring:

Assume a $T$-coloured graph $G\in\cf D$.
We say that $G$ is {\em consistent with its
$T$-colouring} (shortly {\em consistent}) if the following holds:
\begin{itemize}
\item
there exists an induced supergraph $H\supseteq_i G$, $H \in\cf D$ 
and a tree $T_H\in\cf U$ such that $H\simeq T_H^J$,
the isomorphism mapping of $T_H^J\simeq H$ restricted to $\ell(T)$
is the given $T$-colouring of $G$, and
\item
$T\subseteq T_H$ (sharing the root with $T$) and $T_H$ reduces to $T$.
\end{itemize}

The definition automatically gives that the property of consistency with
$T$ is hereditary (closed under induced colour-preserving subgraphs)
on the universe of $T$-coloured graphs from $\cf D$.
Therefore, by Theorem~\ref{thm:Ding} and Proposition~\ref{prop:obstacles}, 
there is a finite set $Obst(T)$ of $T$-coloured
graphs such that $T$-coloured $G$ is consistent if and only if $G$ has no
induced subgraph isomorphic to a member of $Obst(T)$.
Consequently, this property can be expressed by an \FO formula
$\pconsistent T$.

We also use one special property of the interpretation $J$
from Lemma~\ref{lem:tdinterpret},
that $J$ is ``tree-ordered'', meaning that an edge $uv$ 
in the graph $T^J$ may exist only if the interpretation of $u$ in $T$ is an
ancestor of that of $v$ or vice versa.
This property, informally, will allow us to simply identify limbs of $T$
within $T^J$ using connectivity:
if $T_1,T_2$ are two disjoint limbs in $T$, then there is no edge between
$\ell(T_1)$ and $\ell(T_2)$ in the graph $T^J$.

Now we use the previous to describe the property that $G\in\cf D$
{\em reduces to} $U\in\cf U_0$, i.e., that there exists $T_G\in\cf U$ such that 
$G\simeq T_G^{\>J}$ and $T_G$ reduces to $U$.
(Recall that if $G$ reduces to $U$, then $G\models\psi\iff U\models\psi^J$.)
Considering the restriction of the mapping $T_G^{\>J}\simeq G$ to $\ell(U)$
we say that $G$ reduces to $U$ {\em respecting this $U$-colouring} of~$G$.
We introduce the following shorthand notation:
\begin{itemize}
\item
For a tree $T\in\cf U$, let $\widehat{x_{T}}$ denote a collection of variables
indexed by the elements of $\ell(T)$, i.e., $\widehat{x_{T}}=\big(x_v:
v\in\ell(T)\big)$.
Let $\widehat{x_{T}}\cup\widehat{x_{T'}}$ stand for a union of two such
collections and $z\in\widehat{x_{T}}$ mean that $z=x_v$ for some member
$x_v$ of $\widehat{x_{T}}$.
\item
Let $\cf L(T)$ be the set of all non-l-isomorphic limbs $B\subsetneq T$
that reach the considered reduction threshold, i.e.,
if $B$ is of node $v\in V(T)$ and of height~$i$, 
then at least $f(i)-1$ other limbs of $v$ in $T$ are l-isomorphic to~$B$.
For $B\in\cf L(T)$ we denote by $T\odot B$ a tree such that $T\odot
B$ reduces in one step to $T$, i.e., $T\odot B$ results by
adding a disjoint copy of $B$ as a sibling of $B$ into~$T$.

\item
We write $G\models\pconsistent T(\widehat{x_{T}})$ in \FO to mean that $G$ is
consistent with the $T$-colouring given by an assignment of $\widehat{x_{T}}$.
\item
We write $G\models\pconnected(\widehat{x_{T}}, y,z)$ to mean that
$y,z\not\in\widehat{x_{T}}$ and
$y$ and $z$ are connected in $G$ by a path avoiding all $\widehat{x_{T}}$.
Although connectivity is not in \FO in general, our formula $\pconnected$ 
is \FO over $\cf D$ since graphs of bounded tree-depth are hereditary and 
have bounded diameter as well \cite{Sparsity}.
\end{itemize}

\noindent Let $T_1\in\cf U$, and let $T_0$ be a limb of $T_1$ or $T_0=T_1$.
We write in \FO that $G\models\preducex(T_1,T_0)(\widehat{x_{T_1}},y)$
with the following intended (and so far informal) meaning:
There exists a graph $H$ and a tree $T_H\in\cf U$
such that $T_H\supseteq T_1$, and
\begin{itemize}
\item $G\subseteq_i H\simeq T_H^{\>J}$, the elements of $\ell(T_1)$ coincide
with the assignment of $\widehat{x_{T_1}}$ in $G$, and $T_H$ reduces to $T_1$,
\item the connected component of $G-\widehat{x_{T_1}}$ containing $y$
recursively reduces (within the reduction of $T_H$ onto $T_1$) to $T_0$.
\end{itemize}
This is formally written down as follows
\def\BPsi{\mbox{\boldmath$\Psi$}}
\begin{align}
\label{eq:reducex}
\preducex(T_1,T_0)(\widehat{x_{T_1}},y)
	\equiv&
\bigvee_{T_2\in\cf L(T_0)} \exists\, \widehat{z_{T_2}}.
	~~\BPsi_{T_1,T_0,T_2} 
\intertext{where\vspace*{-2ex}}
\BPsi_{T_1,T_0,T_2} 
\equiv& \bigwedge_{v\in V(T_2)} \pconnected(\widehat{x_{T_1}}, y,z_v)
\>\wedge \label{eq:reducex2} \\&\nonumber
	\pconsistent{T_1\odot T_2}(\widehat{x_{T_1}}\cup\widehat{z_{T_2}})
\>\wedge \\\nonumber&
	\>\forall y'\, \big[(
		\pconnected(\widehat{x_{T_1}}, y,y')
	 \wedge y'\not\in \widehat{z_{T_2}})	\,\to
\\&\qquad\label{eq:reduce4}
	\preducex(T_1\odot T_2,T_2)
		(\widehat{x_{T_1}}\cup\widehat{z_{T_2}}, y')
	\big]
,\intertext{except that for $T_2$ of height $0$ (i.e., $T_2$ a single node,
hence singleton $\widehat{z_{T_2}}$!) it is}
\label{eq:reducex0}
\BPsi_{T_1,T_0,T_2} 
	\equiv&~ y\in\widehat{z_{T_2}} \wedge
	\forall y'(\pconnected(\widehat{x_{T_1}}, y,y')\to y'=y) \wedge
\nonumber\\&
	\pconsistent{T_1\odot T_2}\big(\widehat{x_{T_1}}\cup\widehat{z_{T_2}}\big)
.\end{align}

\noindent Note that in \eqref{eq:reducex} the only role of $y$ is to be a
representative of a connected component of $G-\widehat{x_{T_1}}$;
the outcome of $G\models\preducex(T_1,T_0)(\widehat{x_{T_1}},y)$ is
invariant upon the choice of $y$ from the same component.
Task (III) is now finished with
\begin{equation}
\label{eq:reduceto}
\preduceto(U)  \equiv~
	\exists\, \widehat{x_{U}} \,\big\{
	\pconsistent{U}(\widehat{x_{U}}) \wedge
	\,\forall z\, ( z\in \widehat{x_{U}} \vee
	\preducex(U,U)(\widehat{x_{U}}; z))
\big\}
.\quad\end{equation}

\noindent Assume that $G\in\cf D$ reduces to $U\in\cf U_0$.
Then, following the steps of this reduction, it is a simple routine to
verify that $G\models\preduceto(U)$ (one always chooses $H=G$ when speaking
about consistency).
On the other hand, assume \eqref{eq:reduceto} $G\models\preduceto(U)$.
Fix a satisfying assignment of $\widehat{x_{U}}$ into $V(G)$ as the
$U$-colouring of $G$.
If $|\ell(U)|=|V(G)|$, then $G\simeq U^J$ by
$G\models\pconsistent{U}(\widehat{x_{U}})$.
Otherwise, for every connected component $P_0$ of $G-\widehat{x_{U}}$
choose $p\in P_0$; hence \eqref{eq:reduceto} $G\models\preducex(U,U)(\widehat{x_{U}};p)$.
We now analyze the recursive definition \eqref{eq:reducex} of
$\preducex(U,U)$, aiming to show the following claim:
the induced subgraph $G_1=G[\widehat{x_{U}}\cup P_0]$ reduces to $U$
respecting the $U$-colouring induced by $\widehat{x_{U}}$.
If this is true for each component $P_0$ then, clearly, $G$ itself reduces
to $U$.

Let initially $U_0=W_0=U$, and set $W_1=T_2$ where $T_2\in\cf L(W_0)$ of height $>0$ 
is a satisfying branch (any one) of the ``big'' disjunction in \eqref{eq:reducex}
and fix a satisfying assignment of $\widehat{z_{W_1}}$ into $P_0$ thereafter.
Let $U_1=U_0\odot W_1$ and (with a negligible abuse of notation)
$\widehat{x_{U_1}}=\widehat{x_{U_0}}\cup\widehat{z_{W_1}}$.
If $|\ell(W_1)|=|P_0|$, then we are immediately done with the claim by
$\pconsistent{U_1}(\widehat{x_{U_1}})$ in \eqref{eq:reducex2}:
according to the hereditary property of consistency it is $G_1\simeq U_1^J$
and `$U_1\to U_0$' is a valid reduction step.
Otherwise, recursively for every connected component $P_1$ of $G_1-\widehat{x_{U_1}}$
(which is one of components of $G-\widehat{x_{U_1}}$, too)
we choose $q\in P_1$; hence \eqref{eq:reduce4}
$G\models\preducex(U_1,W_1)(\widehat{x_{U_1}};q)$.

In each recursion branch (say, by DFS on the recursion tree)
we continue with the previous argument at ``depth'' $i\geq1$:
we analyze \eqref{eq:reducex}
$G\models\preducex(U_i,W_i)(\widehat{x_{U_i}};q)$ over every connected
component $P_i\ni q$ of $G_i-\widehat{x_{U_i}}$, 
denote by $G_{i+1}=G[\widehat{x_{U_i}}\cup P_i]$,
and analogously obtain $W_{i+1}$ ($=T_2\in\cf L(W_i)$ from a satisfying
branch) and $U_{i+1}=U_i\odot W_{i+1}$, with an assignment of
$\widehat{z_{W_{i+1}}}$ into $P_i$.
This recursion is finite since the height of $W_{i+1}$ strictly decreases
and then generally terminates at \eqref{eq:reducex0} for $W_{i+1}$ of height~$0$.
In this terminal case we get by \eqref{eq:reducex0}
for $\BPsi_{U_i,W_i,W_{i+1}}$ that the assignment of               
$\widehat{z_{W_{i+1}}}$ is $q$ and $\{q\}=P_i$,
and so $U_{i+1}^J\simeq G_{i+1}$ by $\pconsistent{U_{i+1}}$.
This finishes the proof since $U_{i+1}$, by the construction, reduces to~$U$.
\end{proof}

Notice how much of the previous proof is very general, using just the fact
that the graph class $\cf D$ has {\em some} hereditary    
\MSOi interpretation $J$ in the class $\cf U$.
Interestingly, well-quasi-ordering of $\cf D$ under induced subgraphs can be
derived already from this assumption and Theorem~\ref{thm:Ding}.
There is, however, one technical point in the proof which heavily depends on special
properties of the interpretation $J$---it is the use of predicate
$\pconnected(\widehat{x_{T}},y,z)$ in the proof of Theorem~\ref{thm:EGT}.
This is based on a quite unique property of tree-depth,
and before attempting to give a similar result for other graph classes
interpretable in $\cf U$, we have to find a suitable replacement property
in the next section.

\subsection{Tree-models and their properties}
\label{sec:treemodels}

Graph classes of bounded shrub-depth have been introduced recently in
\cite{GHNOOR12} as those having \MSOi interpretations in
the class(es) of rooted labelled trees of fixed height.
Equivalently, they are defined by a very special kind of
a hereditary interpretation:

\begin{defi}[Tree-model \cite{GHNOOR12}]
\label{def:tree-model}
We say that a graph $G$ has a {\em tree-model of $m$ colours and depth $d$}
if there exists a rooted tree $T$ (of height $d$) such that 
\begin{enumerate}[label=\roman*.]\vspace*{-2pt}
\item the set of leaves of $T$ is exactly $V(G)$,
\item the length of each root-to-leaf path in $T$ is exactly~$d$,
\item each leaf of $T$ is assigned one of $m$ colours
(\,$T$ is {\em $m$-coloured}), 
\item\label{it:tree-model-edge}
and the existence of a $G$-edge between $u,v\in V(G)$ depends solely
on the colours of $u,v$ and the distance between $u,v$ in $T$.
\end{enumerate}\vspace*{-2pt}
The class of all graphs having such a tree-model is denoted by $\TM dm$.
\end{defi}

\begin{figure}[tb]
\begin{center}
\scalebox{1}[-1]{\includegraphics[width=.5\hsize]{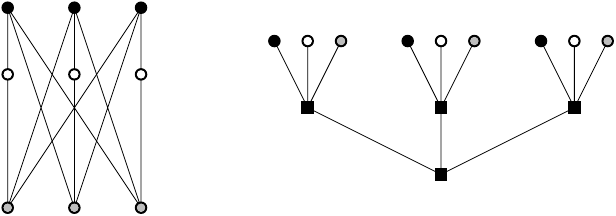}}
\end{center}
\caption{The graph obtained from $K_{3,3}$ by subdividing a matching belongs to
$\TM 23$.}
\label{fig:TMillustration}
\end{figure}
Notice that, in (\ref{it:tree-model-edge}.) of the definition,
the existence of an edge $uv$ depends on finite information.
For example, $K_n\in\TM11$ or $K_{n,n}\in\TM12$.
Definition~\ref{def:tree-model} is further illustrated in
Figure~\ref{fig:TMillustration}.
It is easy to see that each class $\TM dm$ is closed under complements 
and induced subgraphs, 
but neither under disjoint unions, nor under subgraphs.

\begin{defi}[Shrub-depth \cite{GHNOOR12}]
\label{def:shrub-depth}
A class of graphs $\cf S$ has {\em shrub-depth} $d$
if there exists $m$ such that $\cf S\subseteq\TM dm$,
while for all natural $m$ it is $\cf S\not\subseteq\TM{d-1}m$.
\end{defi}

Note that Definition~\ref{def:shrub-depth} is asymptotic as it makes sense 
only for infinite graph classes;
the shrub-depth of a single finite graph is always at most one
($0$ for empty or one-vertex graphs).
For instance, the class of all cliques has shrub-depth $1$.
Similarly, although Definition~\ref{def:tree-model} does not explicitly
specify rules for the existence of edges (\ref{it:tree-model-edge}.),
Definition~\ref{def:shrub-depth} suggests a natural associated \FO
interpretation $J$ of the class $\cf S$:
\begin{defi}[Shrub interpretation]
\label{def:shrubinterpret}
For a graph class $\cf S$ of shrub-depth~$d$, a
{\em shrub interpretation} of $\cf S$ in the class $\cf U$             
of rooted labelled trees of height $d$ is a hereditary
interpretation $J$ satisfying the following:
for $G\in\cf S$ with a tree-model $T$ of depth $d$, it is 
$G\simeq T_1^J$ where $T_1$ inherits $T$ with all the leaf colours,
and each leaf $v$ of $T_1$ is additionally equipped with all labels of the form
$(i,c)$ where $c$ is the $T$-colour of a vertex $u$ adjacent to $v$ in $G$ 
such that the distance between $u,v$ in $T$ is $2i$.
\end{defi}
\begin{lem}
\label{lem:shrubint}
A shrub interpretation is \FO definable for each fixed~$d$.
\qed\end{lem}

For more relations of shrub-depth to other established concepts
such as cographs or clique-width we refer the reader to \cite{GHNOOR12}.
Here we just summarize:

\begin{prop}[\cite{GHNOOR12}]
\label{pro:td-shrub-cw}
Let $\cf G$ be a graph class and $d$ an integer. Then:
\begin{enumerate}[label=\alph*)]
\item
If $\cf G$ is of tree-depth $\leq d$, then $\cf G$ is of shrub-depth~$\leq d$.
\item
If $\cf G$ is of bounded shrub-depth, then $\cf G$ is of bounded
clique-width.\qed
\end{enumerate}
\end{prop}

\begin{prop}(See also \cite{GHNOOR12} for a more general statement.)
\label{pro:treemodelWQO}
Let $\cf S$ be a graph class of bounded shrub-depth.
Assume the graphs of $\cf S$ are arbitrarily coloured from a finite set of colours.
Then $\cf S$ is well-quasi-ordered under the colour-preserving induced subgraph order.
\end{prop}
\begin{proof}
Consider an infinite sequence $(G_1,G_2,\dots)\subseteq\cf S$, and the
corresponding tree-models $(T_1,T_2,\dots)$.
Let $T_i^+$, $i=1,2,\dots$, denote the rooted tree with leaf labels composed
of the colours of $T_i$ and the colours in $G_i$.
By Theorem~\ref{thm:Ding},
$T_1^+,T_2^+,\dots$ of bounded diameter are WQO under 
rooted coloured subtree relation, and, consequently, so are
the coloured graphs $G_1,G_2,\dots$, as desired.
\end{proof}

For the rest of this section we will focus on the shrub interpretation $J$
associated with a graph class $\cf S$ of bounded shrub-depth.
This allows us to smoothly adapt the notions introduced during the proof of
Theorem~\ref{thm:EGT}.
In particular, $\ell(T)$ (the domain of $T^J$) is now the set of leaves of $T$
and the notion of a $T$-colouring corresponds to that.
We also literally adopt the definition of $G\in\cf S$ being
{\em consistent with its $T$-colouring}.
Thanks to Proposition~\ref{pro:treemodelWQO}, this property can be expressed
by an \FO formula $\pconsistent T$ (depending on $\cf S$), too.

Though, we need a bit stronger and crucial property of separability for $T$.
It deals with a {\em$(T,red,blue)$-coloured} graph
which is a $T$-coloured graph in which two other (non-$T$-coloured)
vertices are assigned colours $red$ and $blue$.
When the $red$ vertex is $x$ and the $blue$ one is $y$ in a graph $H$, 
then we call this also a {\em$(T,x,y)$-colouring} of~$H$.

\begin{defi}[Separability for $(T,red,blue)$-coloured]
\label{def:separability}%
Let $\cf S$ be a class of bounded shrub-depth with a shrub interpretation
$J$ in $\cf U$ (cf.~Definition~\ref{def:shrubinterpret}).
Assume a $(T,red,blue)$-coloured graph $G$ where $r,b$ denote the 
vertices of colours $red$ and $blue$, respectively.
We say that $G$ is {\em separating} ($r$ from $b$, implicitly) {\em for $T$} if
there exists an induced supergraph $H\supseteq_i G$ and 
a tree $T_H\in\cf U$, $H\simeq T_H^J$,
such that the following hold:
\begin{itemize}
\item $T\subseteq T_H$ (sharing the root with $T$), 
the isomorphism mapping of $T_H^J\simeq H$ restricted to $\ell(T)$
is the given $T$-colouring of $G$, and $T_H$ reduces to $T$;
\item the least common ancestor of the nodes interpreting $r,b$ 
within $T_H$ belongs to~$V(T)$.
\end{itemize}
\end{defi}
Note; separability clearly implies consistency with $T$.
And again by Proposition~\ref{pro:treemodelWQO}, the separability 
property can be expressed with an \FO formula 
$\pseparable T(\widehat{x_{T}}, r,b)$ over the universe
of $(T,r,b)$-coloured graphs from~$\cf S$.

\smallskip

To make practical use of the (generally vague) separability property, we have to restrict
our domain to so called unsplittable tree-models, as follows.

Assume $T$ is a tree-model of a graph $G$, and $B$ is a limb of $v\in V(T)$,
such that $W$ is the set of leaves of $B$.
We say that a tree-model $T'$ is obtained from $T$ by 
{\em splitting $B$ along} $X\subseteq W$ if
a disjoint copy $B'$ of $B$ with the same parent $v$ is added into $T$, 
and then $B$ is restricted to all its root-to-leaf paths ending in $W\sem X$
while $B'$ is restricted to all such paths ending in $X'$ 
(the corresponding copy of~$X$).
A tree-model $T$ is {\em splittable} if some limb in $T$ can be split along
some subset $X$, making a tree-model $T'$ which represents the same
graph $G$ as $T$ does.
A tree-model is {\em unsplittable} if it is not splittable.
Notice that any tree-model can be turned into an unsplittable one;
simply since the splitting process {\em must end} eventually.

An {\em unsplittable shrub interpretation} $J$ is an interpretation
satisfying Definition~\ref{def:shrubinterpret} with the target class
$\cf U$ containing only trees of unsplittable tree-models.
We also implicitly assume that the threshold function $f$ in the definition
of `$f$-reduce' is always at least~$2$.
Then we claim:

\begin{lem}
\label{lem:separability}
Assume an {\em unsplittable} shrub interpretation $J$ in $\cf U$.
For a graph $G\simeq T_G^J$ let $T_G\in\cf U$ reduce to $T_0$,
and associate $G$ with this $T_0$-colouring.
Let $L_1,\dots,L_k$ denote the leaf sets of all $k$ subtrees in $T_G-V(T_0)$.
If $\sim$ is the binary relation on $V(G)\sem\ell(T_0)$ defined as
$x\sim y$ iff the corresponding $(T_0,x,y)$-colouring of $G$ is {\em not} separating,
then $\sim$ is a relation of equivalence whose
classes coincide with the sets $L_1,\dots,L_k$.
\end{lem}

Before giving a proof, we add the following immediate corollary which
may be interesting on its own (though not used here).
\begin{cor}\label{cor:separability}
If $J$ is unsplittable, then the partition $L_1,\dots,L_k$ from
Lemma~\ref{lem:separability} is unique for given $G$ and $T_0$ (regardless of $T_G$).
\qed\end{cor}
\begin{rem}
Note that
it is {\em not possible} to simply relate which equivalence class of $\sim$
corresponds to which subtree of $T_G-V(T_0)$.
This is since the definition of separability deals with a
supergraph $H\supseteq_i G$
and then a tree $T_H\supsetneq T_G$ may reduce to
$T_0$ in a different way than $T_G$ does.
\end{rem}

\begin{proof}[Proof of Lemma~\ref{lem:separability}]
Suppose that $x,y$ belong to distinct ones of the sets $L_1,\dots,L_k$.
Then already $H=G$ in Definition~\ref{def:separability} witnesses 
that the $(T_0,x,y)$-coloured graph $G$ is
separating, and so $x\not\sim y$.

In the other direction, we aim for a contradiction.
Two vertices $p,q$ are called {\em twins} with respect to a set $R$ if the
neighbourhood of $p$ in $R$ equals that of $q$ in~$R$.
Moreover, $P,Q\subseteq V(G)$ are {\em twin sets} in $G$ if there is a
bijection $h:P\to Q$ such that $h$ is an isomorphism of $G[P]$ onto $G[Q]$,
and $p$, $h(p)$ are twins wrt.\ $V(G)\sem(P\cup Q)$ for each~$p\in P$.

We assume (up to symmetry) that $x,y\in L_1$, but
the $(T_0,x,y)$-coloured graph $G$ is separating for~$T_0$.
Hence there is a graph $H\supseteq_i G$ and a tree $T_H\supseteq T_0$,
$H\simeq T_H^J$, such that
the least common ancestor of $x,y$ within $T_H$ belongs to~$V(T_0)$.
If $K_1,\dots,K_{k'}$ denote the sets of leaves of all $k'$ subtrees in
$T_H-V(T_0)$ then, say, $x\in K_1$ but $y\not\in K_1$.
We summarize what this means according to our definitions:
\begin{enumerate}[label=\roman*.]\parskip2pt
\item\label{it:LMM}
Since $T_G$ reduces to $T_0$, the set $L_1$ is the leaf set of a subtree 
$U_1$ in $T_G-V(T_0)$, and this $U_1$ is reducible to $U_1'$ such that
$U_1'$ is isomorphic to (at least) two disjoint sibling limbs $B,B'$ in $T_0$.
Clearly, we may assume that $x,y$ are in $U_1'$.
We consider the induced subgraph $G'=G-(V(U_1)\sem V(U_1'))$, and
$L_1',M,M'\subseteq V(G')$ the leaf sets of $U_1',B,B'$, respectively.
Then $L_1',M,M'$ are pairwise twin sets in~$G'$
by Definition~\ref{def:tree-model}.

\item\label{it:KNN}
The same as in (\ref{it:LMM}.) can be claimed for $H$, $T_H$, and $K_1$ 
in place of $G$, $T_H$, and~$L_1$; giving us sets (pairwise twin) 
$K_1'\subseteq K_1$ and $N,N'\subseteq \ell(T_0)$
in a suitable induced subgraph of~$H$.
We may similarly assume $x\in K_1'$.
Moreover, since $N,N'$ are the leaf sets of some limbs in $T_0$, too,
each of $N,N'$ may either be disjoint or in an inclusion with each of $M,M'$.
Consequently, up to symmetry, $N$ can be assumed disjoint from $M'$.
In the graph $G'$ we, in particular, have that for every vertex
$u\in K_1'\cap V(G')$ there is $w\in N$ such that 
$u,w$ are twins wrt.\ $(L_1'\cup M\cup M')\sem(K_1'\cup N)$.
\end{enumerate}
Now, we get a contradiction (to being unsplittable)
by applying next Lemma~\ref{lem:splittable} with $G:=G'$,
$W:=M,\,W':=M',Y:=L_1'$, $X:=N$, $Z:=K_1'\cap V(G')$.
\end{proof}

\begin{figure}[tb]
\center
\begin{tikzpicture}[scale=1]
    \fill[color=black] (6.0,3) circle (0.33mm);
    \fill[color=gray] (6.55,3) circle (1.9mm);
    \fill[color=gray] (4.15,3) circle (1.9mm);
    \fill[color=gray] (2.55,3) circle (1.9mm);
    \fill[color=gray] (0,3) circle (1.9mm);
\tikzstyle{every node}=[draw, shape=circle, minimum size=0pt,inner sep=0pt, color=black]
    \draw[color=black] (0.16,3) ellipse (4mm and 3mm);
    \draw (2.2,3) ellipse (6mm and 3mm);
    \draw (3.8,3) ellipse (6mm and 3mm);
    \draw[color=black,thick] (6.2,3) ellipse (6mm and 3mm);
    \draw[color=black,thick] (6.7,3) ellipse (4mm and 3mm);
    \draw (0.66,2.4) node (x) [label=left:$\underline{\,G_0}:\qquad X~$] {};
    \draw (2.4,2.4) node (w1) [label=left:$W$] {};
    \draw (3.6,2.4) node (w2) [label=right:$W'$] {};
    \draw (4.2,2.4) node (w3) {};
    \draw (6,2.4) node (y) [label=right:$Y$] {};
    \draw (6.7,2.4) node (z) [label=right:$~Z\quad:\underline{G-G_0}$] {};
\tikzstyle{every path}=[draw,dotted,<->]
    \path (w1) -- (w2);
    \path (w3) -- (y);
    \draw (y) arc (300:240:3.6);
    \draw (z) arc (300:240:6.7);
\tikzstyle{every path}=[draw,dashed,color=red,-]
    \draw (0,3.8) node (xa) [label=below:{\color{red}?}] {};
    \path (1.5,3) -- (2.2,4) node (wa) {} -- (2.9,3);
    \path (3.1,3) -- (3.8,4) node (wb) {} -- (4.5,3);
    \path (wa) -- (3,4.6) node (wc) {} -- (wb);
    \path (wc) -- (2.4,5);
    \path (xa) -- (0.6,5) node [label=right:$\color{red}\qquad T_0$] {};
    \path (5,1.5) -- (5,5);
\end{tikzpicture}
\caption{A situation which cannot happen, in a graph $G$ with an unsplittable
	tree-model $T_0$ of an induced subgraph $G_0\subseteq G$,
	and with the sets $W,W',Y$ and $X,Z$ as in Lemma~\ref{lem:splittable}.}
\label{fig:splittable}
\end{figure}

\begin{lem}
\label{lem:splittable}
Let $G$ be a graph, and $T_0$ be a tree-model of an induced subgraph
$G_0\subseteq_i G$.
Let $T_0$ contain two (disjoint) isomorphic limbs $B,B'$ of a node $v$, and
$W,W'\subseteq V(G_0)$ be the sets of leaves of $B,B'$, respectively.
Let $X\subseteq V(G_0)$ be a set disjoint from $W'$.
If there exist sets $Y,Z\subseteq V(G)\sem V(G_0)$ such that
\begin{enumerate}[label=\alph*)]
\item
$W,W',Y$ are pairwise twin sets in $G_0$,
\item
for each $z\in Z$ there is some $h(z)\in X$ such that
$z,h(z)$ are twins with respect to \mbox{$(W\cup W'\cup Y)\sem(Z\cup X)$,}
\item
$Y\not=Y\cap Z\not=\emptyset$,
\end{enumerate}
then the tree model $T_0$ is {\em splittable}.
\end{lem}

The statement is illustrated in Figure~\ref{fig:splittable}.

\begin{proof}
Our aim is to prove that, assuming a)\,--\,c) to be true,
the limb $B$ in $T_0$ can be split along the set $W_Z\subseteq W$ which in the twin set
relation (a) corresponds to $Y\cap Z$ (and analogously for $B'$).
We take arbitrary $z\in Y\cap Z$ and $y\in Y\sem Z$,
and denote by $w,t\in W$ and $w',t'\in W'$ the vertices such that
$w,w'$ correspond to $y$ and $t,t'$ to $z$ in the twin set relation
(in particular, $t\in W_Z$).
By Definition~\ref{def:tree-model}, only the edges between $W_Z$ and 
$W\sem W_Z$ in $G_0$ are potentially affected by the splitting operation on $B$,
and so it is enough to show that 
$wt\in E(G_0)$ iff $w't\in E(G_0)$ to get the desired conclusion.

Assume $wt\in E(G_0)$. Then $yz\in E(G)$ by (a) for the twin pair $W,Y$.
Since $y\not\in Z\cup X$, from (b) and $yz\in E(G)$ we get $yh(z)\in E(G)$.
Then, $h(z)\in X$ implies $h(z)\not\in W'$ which is disjoint from $X$,
and so from (a) for the twin pair $W',Y$ it follows $w'h(z)\in E(G)$.
Furthermore, $w'\not\in Z\cup X$, and hence $w'z\in E(G)$ from (b).
Finally, $w't\in E(G)$ by (a) for the twin pair $W,Y$,
and $w't\in E(G_0)$.
In the exactly same way, $wt\not\in E(G_0)$ implies $w't\not\in E(G_0)$.
\end{proof}

\subsection{Case of bounded shrub-depth}
\label{sub:fomsoshrub}

We get to the main new result of Section~\ref{sec:exprpower}:
\begin{thm}
\label{thm:FOMSOnew}
Let $\cf S$ denote any class of graphs of bounded shrub-depth
(Definition~\ref{def:shrub-depth}).
Then \FO and \MSOi have the same expressive power on $\cf S$.
\end{thm}

\begin{proof}
Our proof follows the same steps as that of Theorem~\ref{thm:EGT}.
\begin{enumerate}[label=(\Roman*)]
\item
Let $J$ be a shrub interpretation (Lemma~\ref{lem:shrubint})
of $\cf S$ in the class $\cf U$ 
of rooted labelled trees of height $\leq d$, for some integer constant~$d$.
By implicitly restricting $\cf U$ we may assume it is unsplittable.
By Theorem~\ref{thm:dtreekern}, there is a finite set
$\cf U_0\subseteq \cf U$ of reduced kernels;
every $T\in \cf U$ reduces to an ``easily definable'' $T_0\in\cf U_0$
such that $T\models\psi^J\iff T_0\models\psi^J$.
\item
For $G\in\cf S$, hence, $G\models\psi$ is equivalent to saying
that $G\simeq T_G^{\>J}$ for a tree $T_G\in\cf U$ such that
$T_G$ reduces to a tree in
 $\cf U_0^{\psi,J}=\{\, U\in \cf U_0: U\models\psi^J \}$ .
\item
It remains to build a desired \FO sentence expressing over $G$ that
(some) implicit unsplittable $T_G$ reduces to a particular unsplittable 
tree $U\in\cf U_0^{\psi,J}$.
\end{enumerate}

\noindent The rest of the proof will again give the details of crucial step (III).
Using the tools developed in Section~\ref{sec:treemodels} this is now a
relatively easy task:
\def\BPsi{\mbox{\boldmath$\Psi$}}%
\begin{align*}
\preducex(T_1,T_0)(\widehat{x_{T_1}},y)
	\equiv&~
	\prebox{sep-is-equivalence}\nolimits_{\,T_1}(\widehat{x_{T_1}})
\>\wedge\> 
\bigvee_{T_2\in\cf L(T_0)} \exists\, \widehat{z_{T_2}}.
	~~\BPsi_{T_1,T_0,T_2} 
\intertext{where\vspace*{-2ex}}
\BPsi_{T_1,T_0,T_2} 
\equiv& \bigwedge_{v\in V(T_2)} \neg\pseparable{T_1}(\widehat{x_{T_1}}, y,z_v)
\>\wedge 
\\&\nonumber
	\pconsistent{T_1\odot T_2}(\widehat{x_{T_1}}\cup\widehat{z_{T_2}})
\>\wedge \\\nonumber&
	\>\forall y'\, \big[(
		\neg\pseparable{T_1}(\widehat{x_{T_1}}, y,y')
	 \wedge y'\not\in \widehat{z_{T_2}})	\,\to
\\&\qquad
	\preducex(T_1\odot T_2,T_2)
		(\widehat{x_{T_1}}\cup\widehat{z_{T_2}}, y')
	\big]
,\intertext{except that for $T_2$ of height $0$ (i.e., $T_2$ a single node,
hence singleton $\widehat{z_{T_2}}$) it is}
\BPsi_{T_1,T_0,T_2} 
	\equiv&~ y\in\widehat{z_{T_2}} \wedge
	\forall y'(\neg\pseparable{T_1}(\widehat{x_{T_1}}, y,y')\to y'=y)\> \wedge
\nonumber\\&
	\pconsistent{T_1\odot T_2}\big(\widehat{x_{T_1}}\cup\widehat{z_{T_2}}\big)
.\end{align*}
Here `$\prebox{sep-is-equivalence}\nolimits_{\,T}(\widehat{x_{T}})$'
asserts that the binary relation defined by
`$\pseparable{T}(\widehat{x_{T}},\cdot,\cdot)$'
is reflexive, symmetric, and transitive on $V(G)\sem\widehat{x_{T}}$,
which has a routine \FO expression.
Task (III) is now finished with
\begin{equation*}
\label{eq:SHreduceto}
\preduceto(U)  \equiv~
	\exists\, \widehat{x_{U}} \,\big\{
	\pconsistent{U}(\widehat{x_{U}}) \wedge
	\,\forall z\, ( z\in \widehat{x_{U}} \vee
	\preducex(U,U)(\widehat{x_{U}}; z))
\big\}
.\quad\end{equation*}

We finally claim that $G\in\cf S$ reduces to $U\in\cf U_0$ if,
and only if, $G\models\preduceto(U)$.
Under Lemma~\ref{lem:separability} this is just a plain repetition of the
arguments in the proof of Theorem~\ref{thm:EGT}.
\end{proof}

\section{Conclusions}

\label{sec:conclus}

Even though our prime motivation was to provide an algorithmic improvement over Courcelle's theorem on
``low-depth'' variants of tree-width and clique-width, the main importance of
our results probably lies in the fact that they provide deeper understanding
of $\MSO$ logic on graph classes interpretable in trees of bounded height. 
This understanding already played crucial role in the proofs
establishing that $\FO$ logic has equal expressive power as $\MSO$ on
these graph classes.

We believe that our results and techniques for obtaining them (especially
the use of well-quasi-ordering) could lead to new results in the future.
Above all we suggest that there is likely no major obstacle to suitable
extensions of the results of Sections~\ref{sec:algoEMSO} and~\ref{sec:exprpower}
to classes of general relational structures.
In particular, the notion of shrub-depth extends easily there.

We conclude with a conjecture stating the converse of our result 
on expressive power or \FO and \MSOi on graph classes of bounded shrub-depth.
\begin{conj}
\label{conj:samepower-shrub}
Consider a hereditary (i.e., closed under induced subgraphs)
graph class~$\cf G$.
If the expressive powers of \FO and \MSOi are equal on $\cf G$,
then the shrub-depth of $\cf G$ is bounded (by a suitable constant).
\end{conj}

Resolving this conjecture would probably require an asymptotic characterization of
shrub-depth in terms of forbidden induced subgraphs.  Such characterization
would probably also allow us to answer the following question: Are there hereditary
graph classes of unbounded shrub-depth having an $\MSOi$ model-checking algorithm
with elementary run-time dependence on the formula?  For tree-depth
and \MSOii the answer is no (unless EXP=NEXP), 
because unbounded tree-depth implies the existence of long
paths and by the result of Lampis~\cite{Lampis-ICALP13} this in turn implies
non-existence of such an algorithm.

\paragraph{\bf Acknowledgements.}
We would like to thank B.~Courcelle for pointing out to us the
``CMSO'' extension of Theorem~\ref{thm:MS1=MS2} contained in \cite{CE12},
and the anonymous referees for their extensive comments
and valuable suggestions on improving the presentation of this paper.

\bibliographystyle{abbrv}
\bibliography{gtbib}

\end{document}